\documentclass[12pt]{elsarticlemod}
\usepackage{lmodern}
\usepackage[T1]{fontenc}
\usepackage[latin1]{inputenc}
\usepackage[a4paper]{geometry}
\usepackage{color}
\usepackage{verbatim}
\usepackage{float}
\usepackage{multirow}
\usepackage{amsbsy}
\usepackage{amstext}
\usepackage{amssymb}
\usepackage{graphicx}
\usepackage{siunitx}
\usepackage[official]{eurosym}
\usepackage{url}
\usepackage{parskip}

\usepackage{amssymb}

\begin{document}

\begin{frontmatter}

\title{Pixel-based tracking detectors for a Low Q$^{2}$ Tagger at EIC - status report}

\author[inst1]{Simon Gardner}
\author[inst1]{Derek I. Glazier}
\author[inst1]{Kenneth Livingston}
\author[inst1]{Dzmitry Maneuski}
\author[inst1]{Daria Sokhan}
\author[inst2]{Jaroslav Adam}

\affiliation[inst1]{organization={School of Physics and Astronomy},
            addressline={University of Glasgow}, 
            city={Glasgow},
            postcode={G12 8QQ}, 
            country={UK}}
\affiliation[inst2]{organization={Faculty of Nuclear Sciences and Physical Engineering},%
            addressline={Czech Technical University in Prague}, 
            city={Prague}, 
            country={Czech Republic}}

\begin{abstract}
The design of pixel-based tracking detectors for a Low Q$^{2}$ Tagger in the Far Backward region of the ePIC detector at EIC is presented. The physics case is outlined, together with estimates of rates and resolutions, and the current design based on Timepix4 technology is introduced. 
\end{abstract}


\end{frontmatter}

%\maketitle

\section{Introduction}
The Electron Ion Collider (EIC) \cite{osti_1765663} will collide electrons with protons and nuclei, with outgoing particles being detected in a primary detector, centred on the interaction point where the beams cross. A Low Q$^{2}$ Tagger is proposed in the \emph{Far Backward} area of the electron Proton/Ion Collider (ePIC) to measure electrons with very low scattering angles which would not be covered by the acceptance of the main detector. To achieve this it will sit close to the beam line at a large distance (>20m) from the interaction point. Early designs were based on a calorimeter with a front layer of scintillating fibres, and focused on integrated flux monitoring, rather than low Q$^{2}$ tagging (where the individual electrons are detected and associated with events causing a trigger in the central detector), which is more challenging, but can offer many advantages for the physics program. Electron tagging has been the core expertise of the Glasgow group for several decades, and our recent role in developing a Low Q$^{2}$ Tagger for CLAS12 \cite{BURKERT2020163419,ACKER2020163475} has given us close insight into the technical and physics issues involved. At the EIC, where the backgrounds rates are very high (particularly from bremsstrahlung), a fully fledged high-rate tracking detector will be required. For example, in a typical 12ns beam pulse there will be multiple tracks originating from the interaction region, together with tracks produced by the synchroton radiation background, and tracks produced via internal rescattering from beamline components. We believe that the current pixel detector technologies (such as Timepix, MAPS, AC-LGAD) will provide a solution, and our strategy has been to develop an initial simulation based on generic pixels, where only size matters, and then, with the addition of improved background event generators, to consider specific pixel detectors. In particular, we have begun by focusing on the newest generation of Timepix device (Timepix4), which combines small pixel size, fast timing, good energy resolution and very high rate capability. The Glasgow group is already involved in the development of new readout and DAQ for Timepix3, and in the construction of a small scale tracker, and we lead the Timepix Work Package within the UK's EIC Infrastructure Consortium \cite{stfcEIC}. As members of the EIC Far Backward Working group we have been closely involved in developing a range of designs for a Low Q$^{2}$ Tagger, and in exploring these using the main Geant4 detector simulation. This document reports on the progress, and also serves as a status report for the forementioned EIC Infrastructure Consortium.

The aims of the work presented here are:
\begin{enumerate}
    \item To consider the physics case for a Low Q$^{2}$ Tagger. Is it well established? Is it worth the investment? What is needed in terms of resolution, event identification etc to make it useful?
    \item \label{generic} To develop a basic Low Q$^{2}$ Tagger design using generic pixel detectors with properties determined by the physics requirements (above) and beam parameters from the EIC design specifications.
    \item \label{specific} To introduce realistic backgrounds, rates and timing into the detector simulation and evaluate the performance of a full tracker design based on Timepix4 as a benchmark device.
    \item \label{other} Compare the performance of other pixel detector options with the Timepix4 solution.
\end{enumerate}

To be clear - this is a report on work in progress, and the current state of things is somewhere between \ref{generic} and \ref{specific} in the list of objectives above (although, these two objectives cannot be completely disentangled). Currently, the physics case is clear, and we have demonstrated that a Low Q$^{2}$ Tagger design using pixel tracker is required, and we are improving the simulation and developing a full design based on Timepix4. 

Here, we briefly outline the physics case for a Low Q$^{2}$ Tagger; give an overview of Timepix, and summarise the work that has been done on design and simulation. We conclude by looking to the future and considering what further effort will be required to move to a fully costed design.

\section{Physics case for a Low $Q^{2}$ Tagger}
The EIC yellow report \cite{khalek2021science} gives many compelling physics cases that would require
a Low $Q^{2}$ Tagger. The primary detector will cover the range of $ -4<\eta<4$ for the measurement of electrons, photons, hadrons, and jets. To cover the kinematically important regions beyond that it requires auxiliary detectors in the far backward and forward regions. Unlike the primary detector 
these regions close to the beamline will have the added complications of high rates and backgrounds
and so any detector technologies used will have to be able to handle this.

The Low $Q^{2}$ Tagger will facilitate measurement of reactions with small cross sections where $Q^{2}$ dependence is not critical. Here we can benefit from the large virtual photon flux for reactions such as : exclusive vector meson production in ep and eA, particularly for upsilon measurements at threshold; a meson spectroscopy programme, particularly with the charmonium-like sector
(XYZ); Time-like Compton scattering; and it will naturally extend the Q2 range of DIS processes.

For Time-like Compton scattering to fully detect the exclusive reaction, we need to detect 
the scattered lepton, recoil proton, and both decay leptons. The most interesting kinematics 
dependence is that of the invariant mass of the two leptons, which will be well reconstructed in 
the primary detector. Measuring the scatted electron will allow the s dependence to be measured as
well as giving some measure of the production four momentum transfer, or t. When coupled with proton 
detection in the far forward region there will be the possibility of applying exclusivity cuts.

For meson photo-production the central detector acceptance is sufficient for all configurations to accept events from $Q^{2} > 0.1$ GeV$^{2}$ to large  $Q^{2}$. 
The lower limit on the scattered electron of $\eta>-3.5$ is restrictive for photo-production events 
in the main detector, especially for higher collision energies. The photo-production of 
Deeply Virtual Meson Production at higher energies will completely depend on
the Low $Q^{2}$ Tagger unless a significant enhancement of the backward region's electron 
acceptance is possible. More acceptance with a Low $Q^{2}$ Tagger would directly
translate into more measured photo-production events. An increased acceptance would benefit $\Upsilon$ photo-production near threshold, where the projected statistical precision is particularly low and the count rates would be dramatically improved in these regions.

The increased virtual photon flux from the tagger would allow a programme of meson spectroscopy to be
undertaken. Of great interest would be the possibility of photo-producing exotic charmonium-like mesons, so-called  XYZ states. Photoproduction is a particularly useful production mechanism as it is capable of, in-principle, generating all possible states with all spins and parities. 
It also removes difficulties in interpretation of peaks coming in fixed mass decays
as kinematic effects, such as triangle singularities, are largely removed. Observation of narrow peaks in invariant masses of final states such as $J/\psi$ and pions should provide conclusive proof of 
the existence of true resonant states.
The Low $Q^{2}$ Tagger would provide the s dependence of the reaction, and if the recoil
proton is detected in the far-forward detectors, provide exclusive final states. Measuring
the scattered electron would also allow an amplitude analysis of the reaction to be performed yielding
the quantum numbers of the states. This will be more powerful if the tagger resolution is 
sufficient to determine the scattering plane and angle, thereby giving the full polarisation state of
the virtual photon. 

\section{\label{timepix}Timepix}
\begin{figure}[H]
\centering{
\includegraphics[clip,height=0.3\columnwidth]{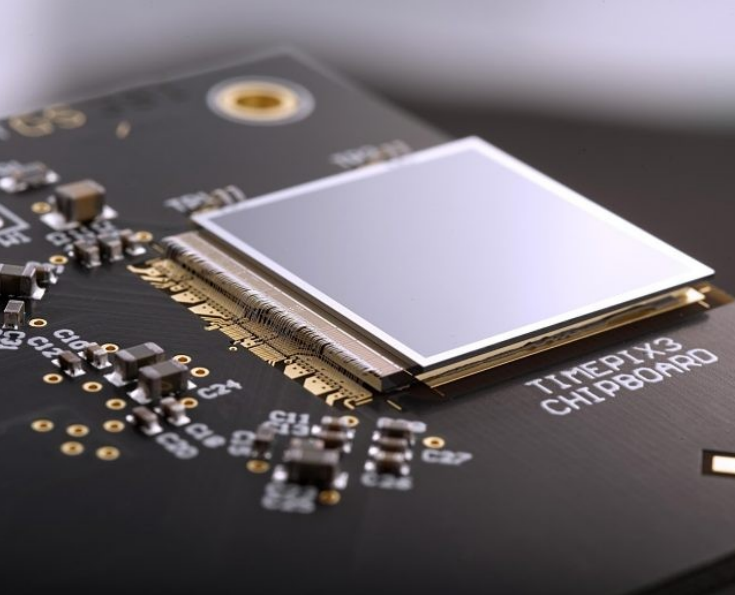}
\hspace{10mm}
\includegraphics[clip,height=0.32\columnwidth]{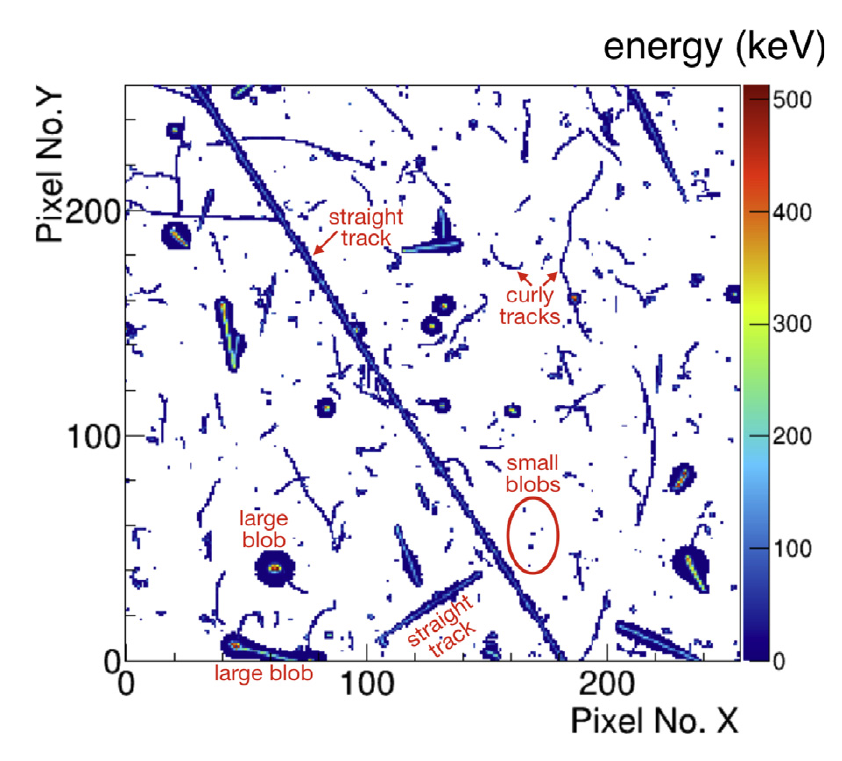}
}
\caption{\label{fig:timepix} Left: Single Timepix3 detector \cite{cernTimepix}. Right: Example of event type identification \cite{Bergmann:2020bdz}}
\end{figure}

Timepix3 is the current generation of hybrid pixel detector (HPD) readout chip from the CERN-based, Medipix collaborations \cite{cernTimepix,Timepix2014,cernMedipix}. It was developed as an extension of imaging sensors, which integrate pixel counts over a time-frame, to a full, data-driven device, where each pixel hit can read be out with time of arrival (ToA) and time over threshold (ToT) information. The chip contains 256 x 256 pixels with a footprint of 14x14 mm, and, in combination with a silicon sensor bonded to the surface, provides excellent rate capability (1.3 kHz / pixel) and  position resolution (< 25 $\mu$m), with timing resolution down to 1.56ns, and very good event identification properties. The further benefit, for applications like the current one, is that Timepix can apply individual thresholds to each pixel, resulting in a very low charged particle detection threshold ($\sim$ 1.5 keV). Figure \ref{fig:timepix} shows a Timepix3 chip with silicon layer mounted on a PCB, and an example of shapes and tracks generated by different event types. Its ancestry is from single-sensor, imaging applications, and the possibilities of building several units into a bigger detector are restricted because one edge must be kept free for bonding. It is \emph{3-side buttable}, so, at most, it can be tiled into a double height row, and therefore has limited use for tracking applications.
The next generation chip, Timepix4 \cite{Xavi2021}, is bigger, better and faster, and with the use of TSV (Through-Silicon-Vias) technology it can be \emph{4-side buttable} with a 99.5\%  active area. It has been developed as a tracking and event identification device, and is ideal for use in high rate, high background environments. Both Timepix3 and Timepix4 are closely related to the VeloPix detector \cite{Poikela_2015} which was developed as the vertex detector (VELO) for the LHCb upgrade. 
At EIC we anticipate that Timepix4 based detectors can have a role in locations where small area, high rate detectors with tracking or position information are required, and good event identification and background rejection is needed. Generally, that means close to the beamline, and specifically, in this instance, as trackers in a Low Q$^{2}$ Tagger. 

The Glasgow University Nuclear and Hadron Physics group started using Timepix3 detectors several years ago to develop a pair spectrometer/polarimeter for tagged bremsstrahlung photons in Mainz \cite{stfcPairPol}. Working in collaboration with the UK's Nuclear Physics Cross Community Support Group (CCG) we have designed a modular readout system based on boards with four Timepix3 hybrids. This is now providing a valuable test and development platform for further Timepix prototyping and design, particularly within the UK's EIC consortium project \cite{stfcEIC}. Here we will complete the 4$\times$Timepix3 module with readout, and will test all aspects of its performance with a view to developing a Low Q$^{2}$ Tagger design based on Timepix4. 

\begin{figure}[H]
\centering{
\includegraphics[clip,width=0.5\columnwidth]{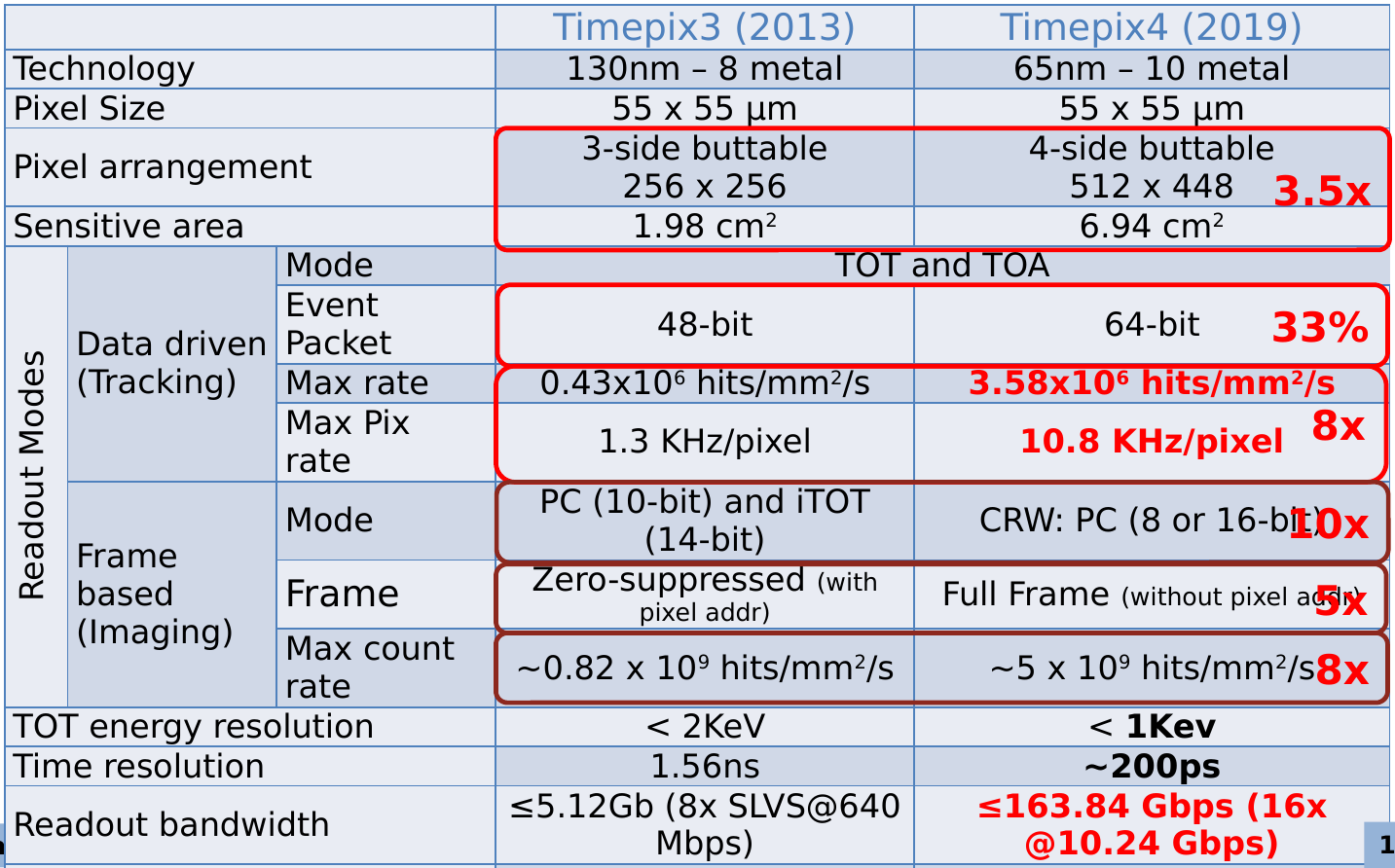}
}
\caption{\label{fig:timepixtable} Comparison of Timepix3 and Timepix4 parameters \cite{Xavi2021}. }
\end{figure}

The details of the electronics and ASIC layout of the Timepix4 chip can be found in refs \cite{Xavi2021,Gromov2021}, and the important parameters from the perspective of particle detection are summarised in Figure \ref{fig:timepixtable}, which compares the performance of Timepix3 and Timepix4. The ToA binning resolution of the chip (195 ps) has improved significantly compared to its predecessor (1.56 ns), and with cluster analysis from ToT (time over threshold) measurements, event timing resolutions significantly lower that this can potentially be achieved, depending on the characteristics of the attached sensor and the type of particle being detected. Cluster analysis can also provide position resolution down to 12.5 $\mu$m, well below the pixel dimension of 55 $\mu$m, over a surface area of 6.94 cm$^2$ (512 x 448 pixels) - a factor of 3.5 increase over Timepix3. This, together with the Timepix4's 4-side \emph{buttability}, provides the opportunity to make a tiled detector with a reasonable area (eg 4x3 chips = 83 cm$^2$), while offering a combination of position resolution, timing resolution, and event identification properties that make it a very strong candidate for the far backward tagger. However, it is the rate capability of Timepix4 which make it ideal for the high intensity, low Q$^2$ region. As is shown in section \ref{sec:rates} this is more that adequate for the rates we anticipate, even in the regions closest to the electron beamline.  

\begin{figure}[H]
\centering{
\includegraphics[clip,width=0.8\columnwidth]{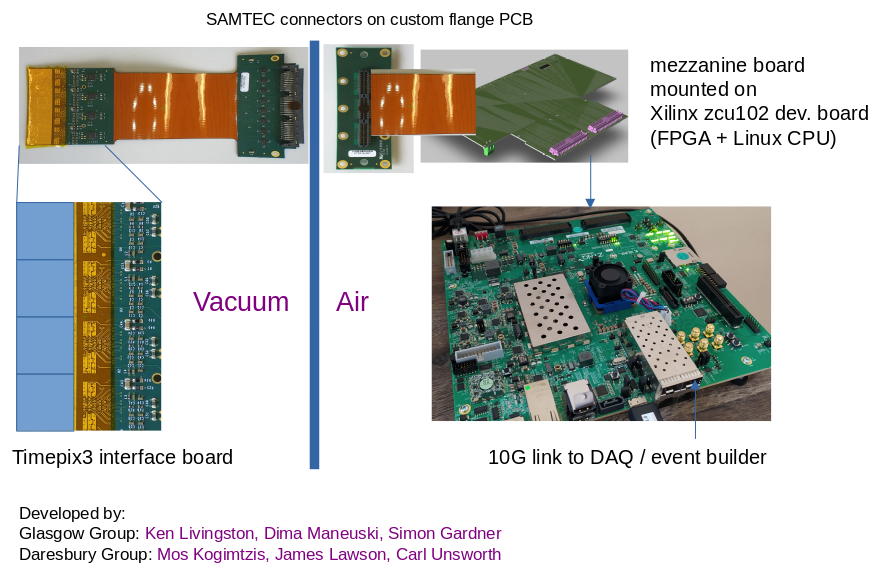}
}
\caption{\label{fig:readout} The Timepix3 readout and DAQ in current development.}
\end{figure}

The current Glasgow / Daresbury Timepix3 module will strongly influence the design of the multi layered Timepix4 detectors for EIC, so an overview of that follows, with some discussion on improvements that will be needed. The components are shown in Figure \ref{fig:readout}, where up to four Timepix3 chips can be bonded onto a PCB and connected to a small flange board with a SAMTEC connector and custom vacuum feedthrough. On the air side this connects to a FPGA Mezzanine Card (FMC) mounted on an FPGA based DAQ/Readout module (Xilinx zcu102 development board), which communicates with the DAQ computer on a 10 Gb optical fibre link. The FPGA board also has a standard (Petalinux) CPU and set of I/O lines which can communicate directly with other identical boards and other detector components. This combination allows a high degree of complex, programmable, event identification and data reduction before sending the data to the DAQ computer over the 10 Gb fibre. The computer will receive data from several readout boards and perform further processing and event identification / rejection before packaging the data to send to the main experiment data stream.

The Timepix3 system development was halted for 18 months due to the COVID pandemic, but progress is back underway. At the time of writing, two Timepix3 chips have been bonded onto interface boards and the readout is being tested and debugged in Daresbury. We anticipate having a working system with two interface boards and silicon sensors by the July 2023, and will carry out tests of hardware, firmware and software before installation in Mainz, where it will be the detector for the new pair spectrometer / polarimeter. 

\begin{figure}[H]
\centering{
\includegraphics[clip,height=0.4\columnwidth]{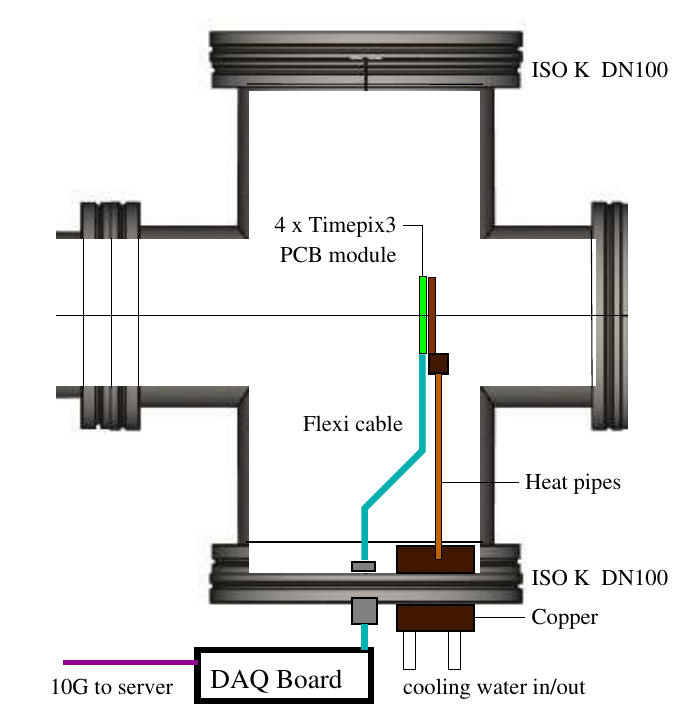}
\includegraphics[clip,height=0.4\columnwidth]{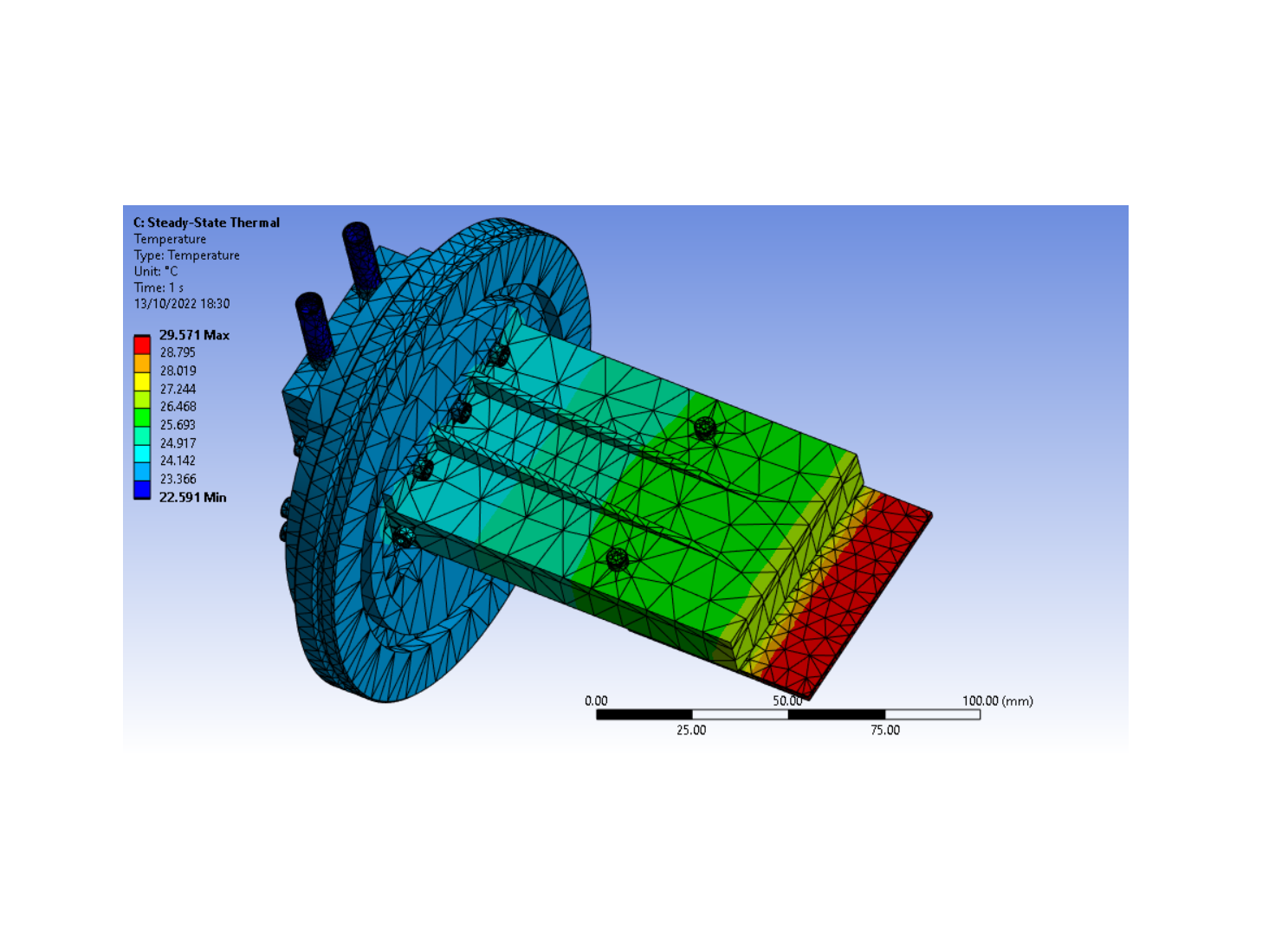}
}
\caption{\label{fig:testrig} Test setup for 4 x Timepix3 modular prototype. Left: schematic of vacuum flange and cross-piece. Right: ANSYS model of flange and cooling.  }
\end{figure}

In parallel with the readout and DAQ development, we have built a portable test rig in the Glasgow labs which holds a  4-unit Timepix3 module on the vacuum side (Figure \ref{fig:readout}) and provides a connection to the DAQ on the outside with a flexible cable. The rig can be seen conceptually in Figure \ref{fig:testrig}, where the essential section is the vacuum cross piece with the Timepix flange; this can be connected locally in a small high vacuum station, but also taken to an electron beam facility (eg MAMI at Mainz) and installed in the beamline for high rate measurements. The system is effectively what would be required for a Low Q$^{2}$ tagger in miniature, allowing us to explore many of the features which will be required for the full scale version. In particular, the cooling requirements will be significant, and we will use the current model to test various approaches to this (Figure \ref{fig:testrig}, right) . At present we use copper heat pipes \cite{heatPipes} together with a temperature controlled chiller mounted on the outside of the flange, in combination with an thermal model implemented with ANSYS.

\begin{figure}[H]
\centering{
\includegraphics[clip,width=0.8\columnwidth]{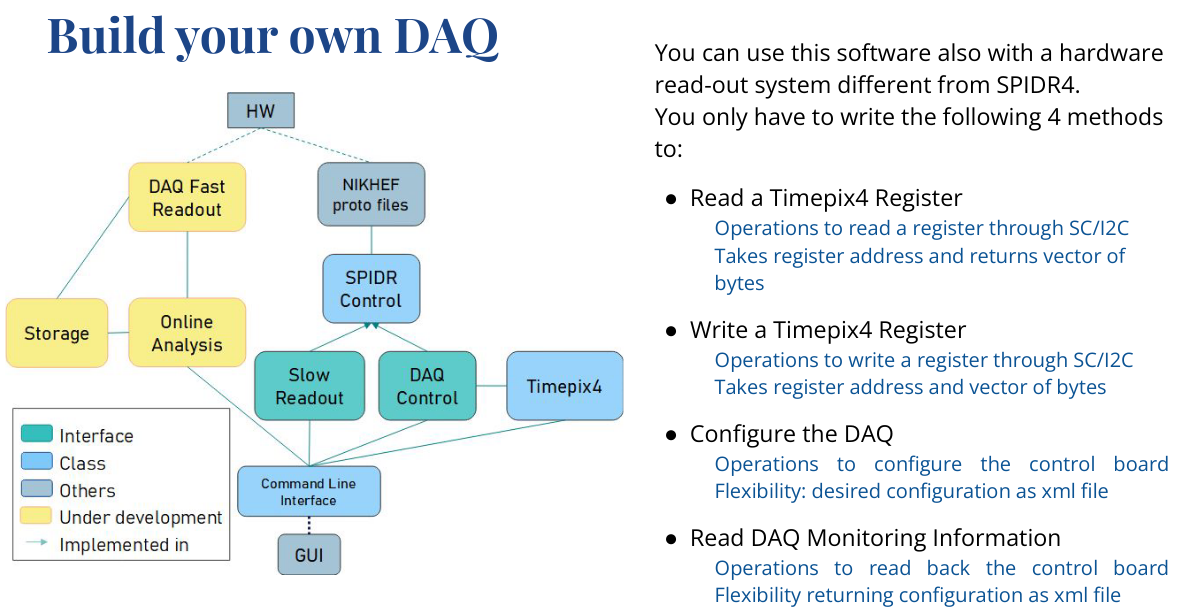}
}
\caption{\label{fig:spidr4} Software toolkit for SPIDR4 and Timepix4 \cite{timepixreadout}}
\end{figure}

The current design was developed for flexible and extendable detector construction, based on single rows of Timepix3 chips, resulting in the basic 4x1 building block show in Figure \ref{fig:readout}, with a DAQ / Readout designed to perform data processing and reduction, and to send/receive signals to up to 16 similar boards. At the outset of the current work we anticipated updating this Timepix3 DAQ/Readout to a 4x4 module of Timepix4 detectors. However, the development of Timepix4, designed with tracking applications in mind, has driven much progress in readout technology, and for EIC we plan to adopt the recently developed SPIDR4 readout\cite{spidr,spidr4_2020}. Figure \ref{fig:spidr4} shows a schematic of the this system, which is conceptually similar to the Daresbury Timepix3 system which we are currently completing, but is an order of magnitude up in terms of data throughput, bandwidth and processing capability. As will be demonstrated in the later section on rate estimates, SPIDR4 already has achieved most of the performance that would be required for EIC. Several different example configuration examples are given in \cite{spidr4_2020}  which show that the current version of this device is already very close to  our needs. 

\begin{figure}[H]
\centering{
\includegraphics[clip,width=0.8\columnwidth]{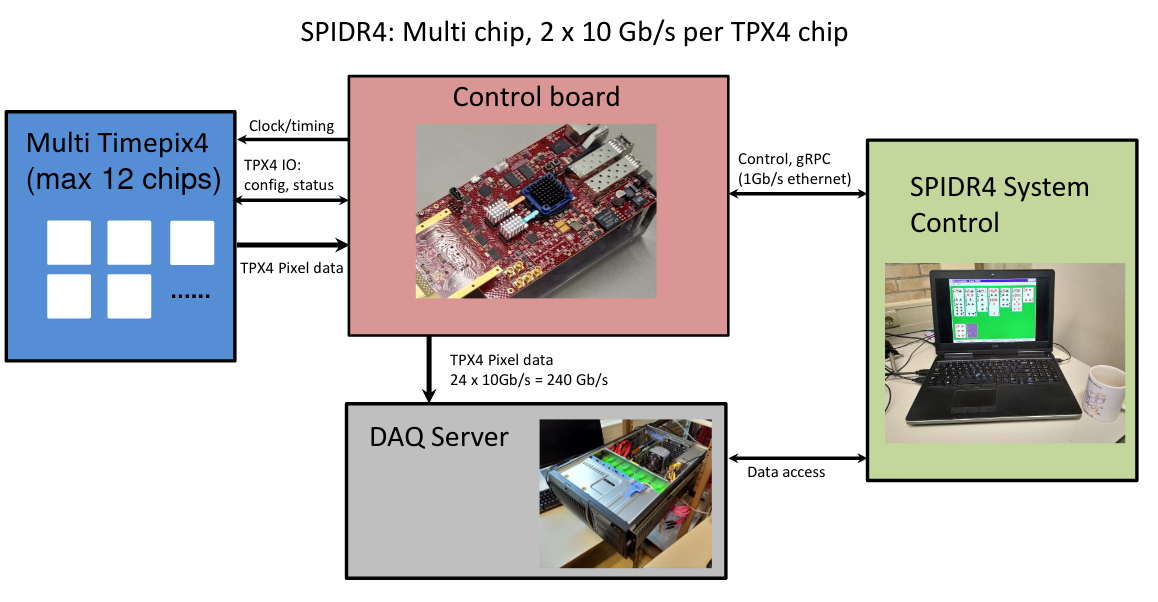}
}
\caption{\label{fig:spidr4daq} The SPIDR4 readout, example configuration \cite{spidr,spidr4_2020}}
\end{figure}

Furthermore, the SPIDR4 technology is well established, and, as it has begun to emerge as the leading solution for Timepix4 applications, it has pushed the community to develop new software tools. In particular, the readout software for Timepix4 from the groups at INFN Ferrara \cite{timepixreadout} is already well advanced and look set to become the de-facto standard for future applications. Figure \ref{fig:spidr4daq} shows how the tools can be harnessed to develop a custom DAQ for a Timepix4-based system such as ours.

In summary, good progress has been made with the Timepix3 readout system. When complete, this will be capable of running medium-size, single-row detectors based on modules with four Timepix3 sensors, resulting in the construction and installation of the prototype pair polarimeter for Mainz \cite{stfcPairPol}. In addition to doing its \emph{day job}, this will also act as a test-bed for modular Timepix detectors and readout, with a view to a future Low Q$^{2}$ Tagger tracker at EIC.

We anticipate using Timepix4 in conjunction with the SPIDR4 DAQ as the basis for the tracking, and, as the development of the Timepix3 detector comes to completion, we have begun to work on a small prototype with two tracker layers (each consisting of a single Timepix4 sensor), and readout using SPIDR4 boards from the next NIHKEF production batch (due in early summer 2023).

The following section presents a simulation of the detector design based on realistic physics and background generation, and generic pixel detectors, and is followed by an evaluation of Timepix4 and other detectors in comparison with requirements extracted from the simulation.

\section{\label{sim}Design and Simulation}
The Low Q$^2$ Tagger along with luminosity detectors make up the far backwards region of the ePIC detector, accepting particles with a polar angle of no more than 10.5~mrad from the electron beam. The Tagger makes use of the drift volume between the first two groups of beamline magnets along the electron beamline after the interaction point. Dipole magnets designed to recirculate the main beam bend electrons of beam energy by 18~ mrad from the experimental axis. Any electrons which have lost energy in an interaction are bent by a larger angle, away from the main beam where they can be detected, tagging the virtual photon of the central interaction. 

The Glasgow group has previous experience designing and analysing data from magnetic spectrometers for fixed target experiments at GlueX, CLAS, A2 Mainz and Bonn where the electrons are separated by energy using a dedicated magnet. The ePIC Low Q$^2$ Tagger comes with an additional challenge as it is parasitic to the recirculating beamline, so scattered electrons pass through a series of fields designed to focus electrons of beam energy. Table \ref{tab:magnets} shows the location and strength of the fields, a scale representation is shown in Figure \ref{fig:farback}.

\begin{table}[H]
    \centering
    \begin{tabular}{l|l|c|c|c|c|c}
        Name        & Type       & Start Pos (m)   & End Pos (m)   & Length (m) & Strength (T)  & Strength (Tm$^{-1}$) \\
        \hline 
        Solenoid    & Solenoid   & 1.82            & -2.02         & 3.84       & 1.7                  &\\
        Q1          & Quadrupole & -5.31           & -7.09         & 1.78       & NA                   & -13.3153\\
        Q2          & Quadrupole & -7.6            & -9            & 1.4        & NA                   & 12.0595\\
        B2A         & Dipole     & -9.61           & -11.39        & 1.78       & 0.192                & NA\\
        B2B         & Dipole     & -11.685         & -14.865       & 3.18       & 0.238                & NA\\
        
    \end{tabular}
    \caption{Magnetic fields between the IP and taggers for an 18~GeV electron beam (Magnetic fields scale linearly with electron beam energy). Positions are relative to the interaction point along the electron beam axis.}
    \label{tab:magnets}
\end{table}

The core aim of the Tagger is to accept as many interaction electrons leaving the beamline dipoles as possible, the design consists of 2 separate tagger stations covering different energy ranges. In order to maximize the coverage at high energy, the second station will be be placed as close to the electron beamline as possible, limited by the requirement to remain outside the profile of 10$\sigma$ of the $e^-$ beam. The parameterization of the beam spread between B2B and Q3 expected based on the interaction conditions and magnets is given by

\begin{equation}
    x = 0.0183z + 44.91\\
\end{equation}

\begin{equation}
    \Delta x = 0.0004207z + 2.492 
\end{equation}
Where this $\Delta$x is the spread perpendicular to x, rather than the beam axis so is fractionally larger than necessary. The design parameters for the beam are shown in Table \ref{tab:beamsettings}.

\begin{table}[h]
    \centering
    \begin{tabular}{l c c c}
         & x & y & z \\
         \hline 
         Beam Divergence [$\mu$rad] & 202 & 187 & \\
         Vertex Spread [mm] & 0.189 & 0.01 & 32.92 \\
    \end{tabular}
    \caption{Design beam parameters for the 18x275 high divergence settings.\cite{beameffects,afterburner}\cite{khalek2021science} (p409)}
    \label{tab:beamsettings}
\end{table}

\begin{figure}[H]
\centering{
\includegraphics[clip,width=0.8\columnwidth]{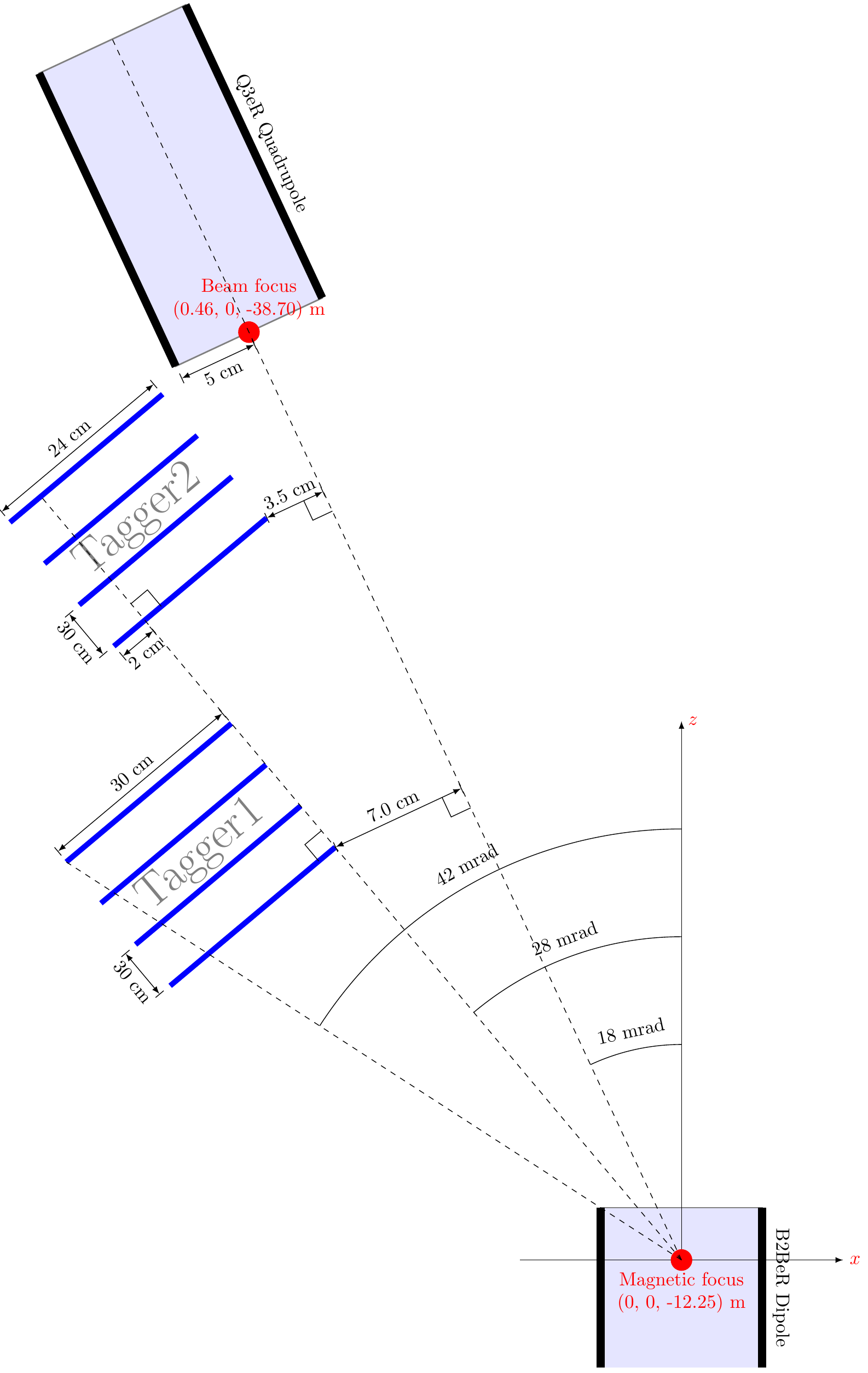}
}
\caption{\label{fig:backschematic} Schematic of the Low-Q2 detector layout in the drift volume between the beamline dipole and quadrupole magnets.}
\end{figure}

Figure \ref{fig:backschematic} shows an overview schematic of the design and constraints. This design has been guided by studies using the ePIC simulation framework discussed further in section \ref{sec:sim}. 

Any electrons bent by more than 41~mrad away from the incoming electron axis in the beamline magnets would end up colliding with the body of the magnet.
  
\subsection{\label{sec:sim}Simulation}

\begin{figure}[H]
\centering{
\includegraphics[clip,width=1.0\columnwidth]{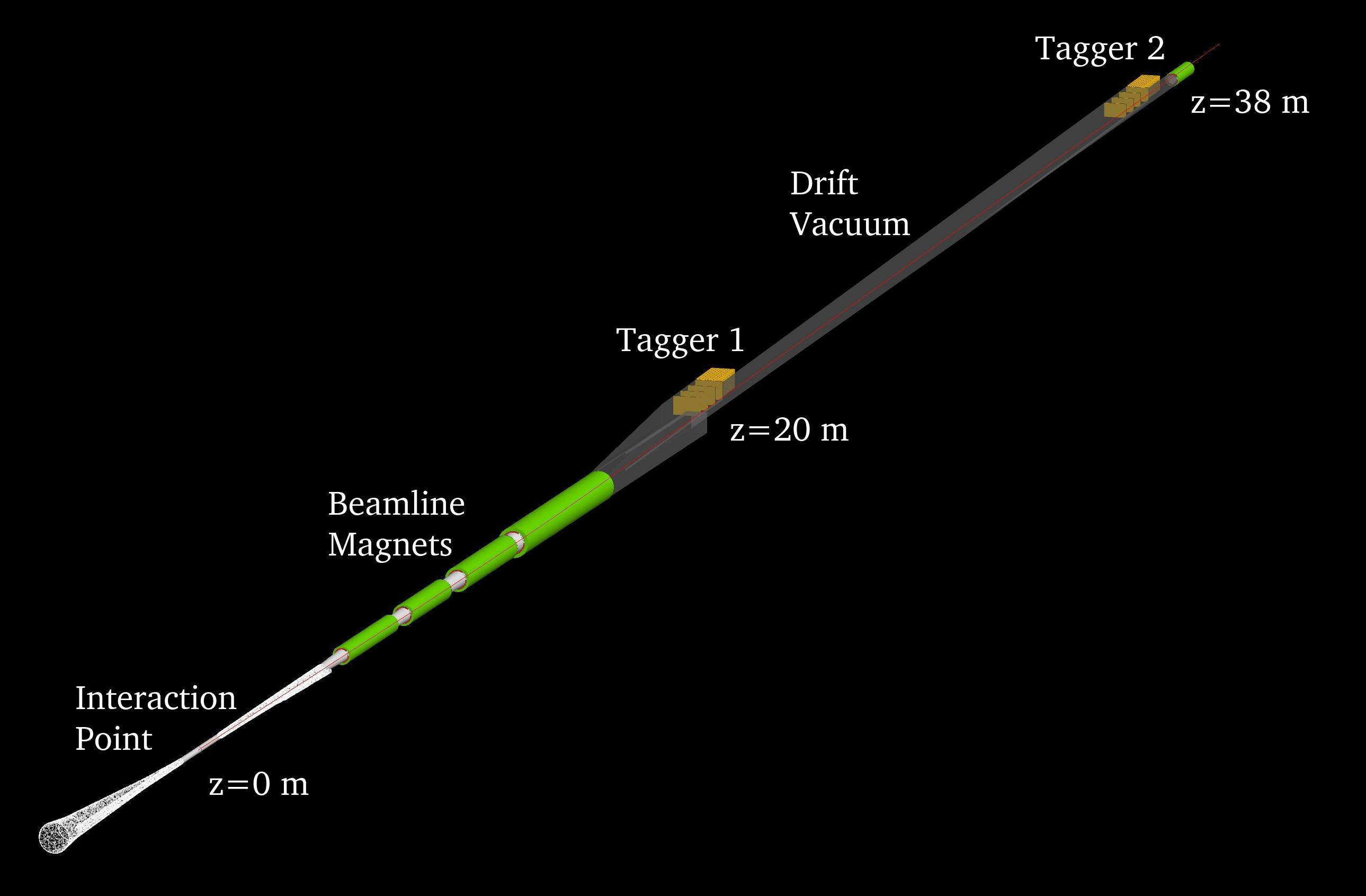}
}
\caption{\label{fig:farback} Far-backwards region of the EIC IP6. All detectors other than the taggers have been removed for clarity.}
\end{figure}

\begin{figure}[H]
\centering{
\includegraphics[clip,width=0.45\columnwidth]{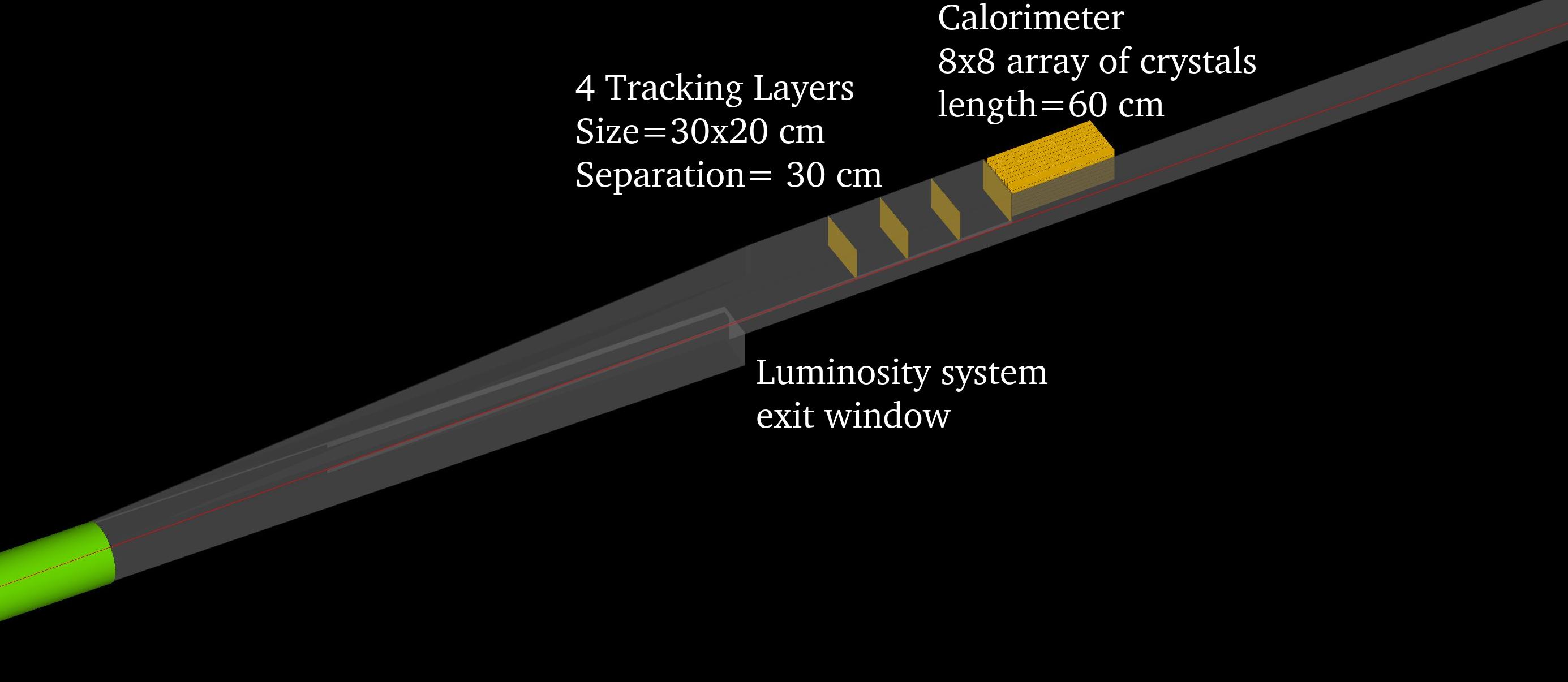}
\includegraphics[clip,width=0.45\columnwidth]{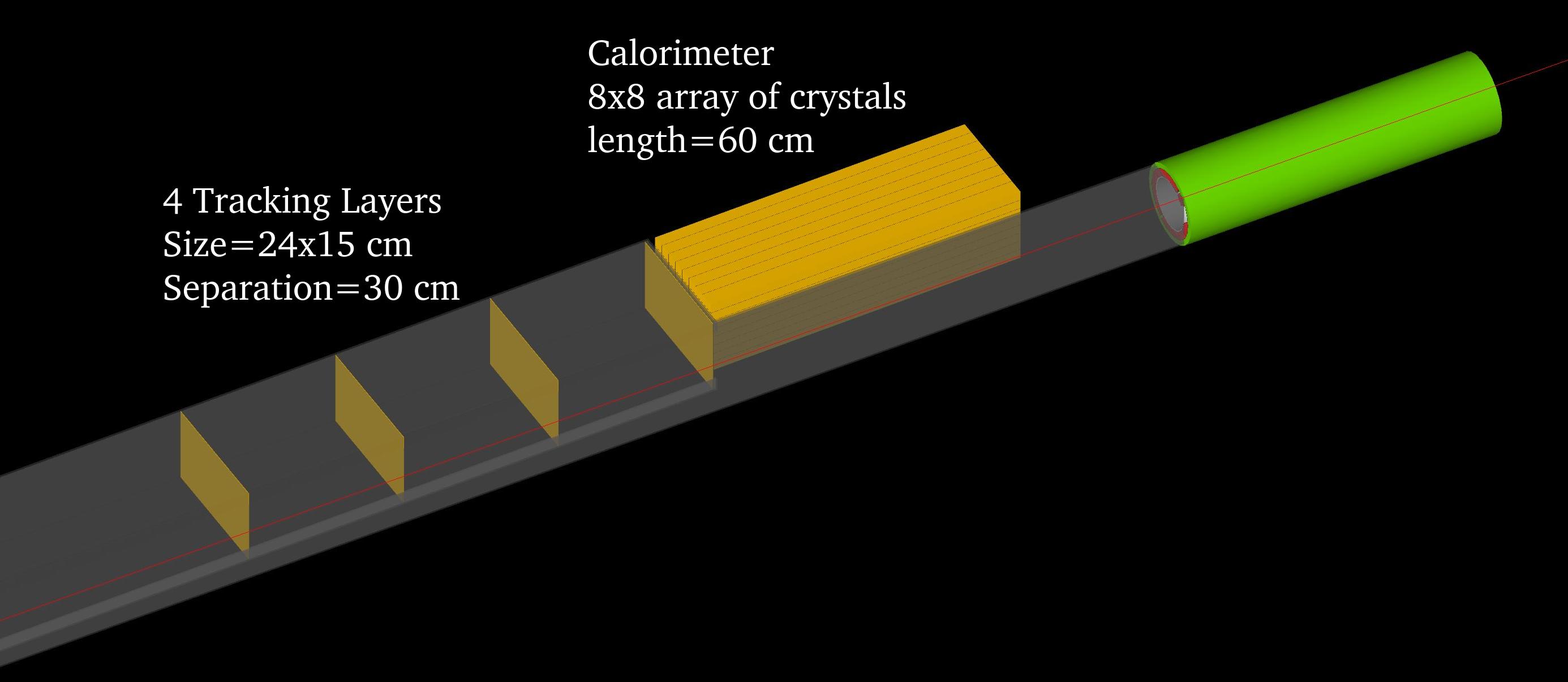}
}
\caption{\label{fig:taggers} Left: Low energy tagger module, green planes are planes of silicon pixel detectors and orange block is an optional calorimeter. Right: High energy tagger module showing an beam electron through the outgoing quadropole magnet.}
\end{figure}

The design was integrated into the ePIC experiment simulation\cite{EPIC_Collaboration_DD4hep_Geometry_Description} in such a way that allowed flexibility for a range of detector studies. The default configuration consisted of two stations of 4 tracking layers placed inside the beamline vacuum and an optional calorimeter (Figures \ref{fig:farback}, \ref{fig:taggers}). Each tracking layer was described as a 400~$\mu$m thick silicon plane with 55~$\mu$m pixel pitch in both x and y.

\subsubsection{\label{sec:eventgen}Event Generator}

Ten million quasi-real photoproduction events and Bremsstrahlung events were generated with the GETaLM event generator\cite{ADAM2022108251}. The studies carried out in this work all use the 18 GeV electron and 275 GeV proton (18x275), "high divergence" beam settings at full design luminosity. The beam parameters relevant to our studies are laid out in Table \ref{tab:beamsettings} these are responsible for some of the hard limits on reconstruction resolutions. The Bremsstrahlung photons generated had a low energy cut of 0.5 GeV in order exclude the peak flux of high energy electrons which are not within the acceptance of the detectors.

These events were run through the simulation to characterise the acceptance, hit distributions and resolution of the taggers, discussed in the next section.

%\subsubsection{\label{sec:response}Detector response}

%A detailed simulation of the Timepix4 detector response from the deposition of energy in the sensor to the digitization of the signal in the chip is supported by the Allpix$^2$ framework\cite{SPANNAGEL2018164} framework. This will allow studies of the detector response to a range of particles with sensors of different materials and thicknesses. Parameterizations of charge sharing distributions and data digitization are essential to create a final model of the detector, taken from these simulations they can be added to the epic simulation/reconstruction. 

%So far no detector response simulations have been conducted for this work. If required, additional, even more in depth simulations of charge collection in various silicon sensor designs such as LGADs will be possible using e.g. Synopsys TCAD.

\subsection{\label{characterisation}Characterisation}
Stand alone code has been developed to analyse the outputs from the ePIC simulation, in the future these characterisation, reconstruction and bench-marking tools will be integrated into the ePIC collaborations event reconstruction software EICrecon.

\subsubsection{\label{sec:acceptance}Acceptance}

\begin{figure}[H]
%\centering{\includegraphics[clip,width=\columnwidth]{EPICAcceptance_18GeV.png}
\centering{\includegraphics[clip,width=\columnwidth]{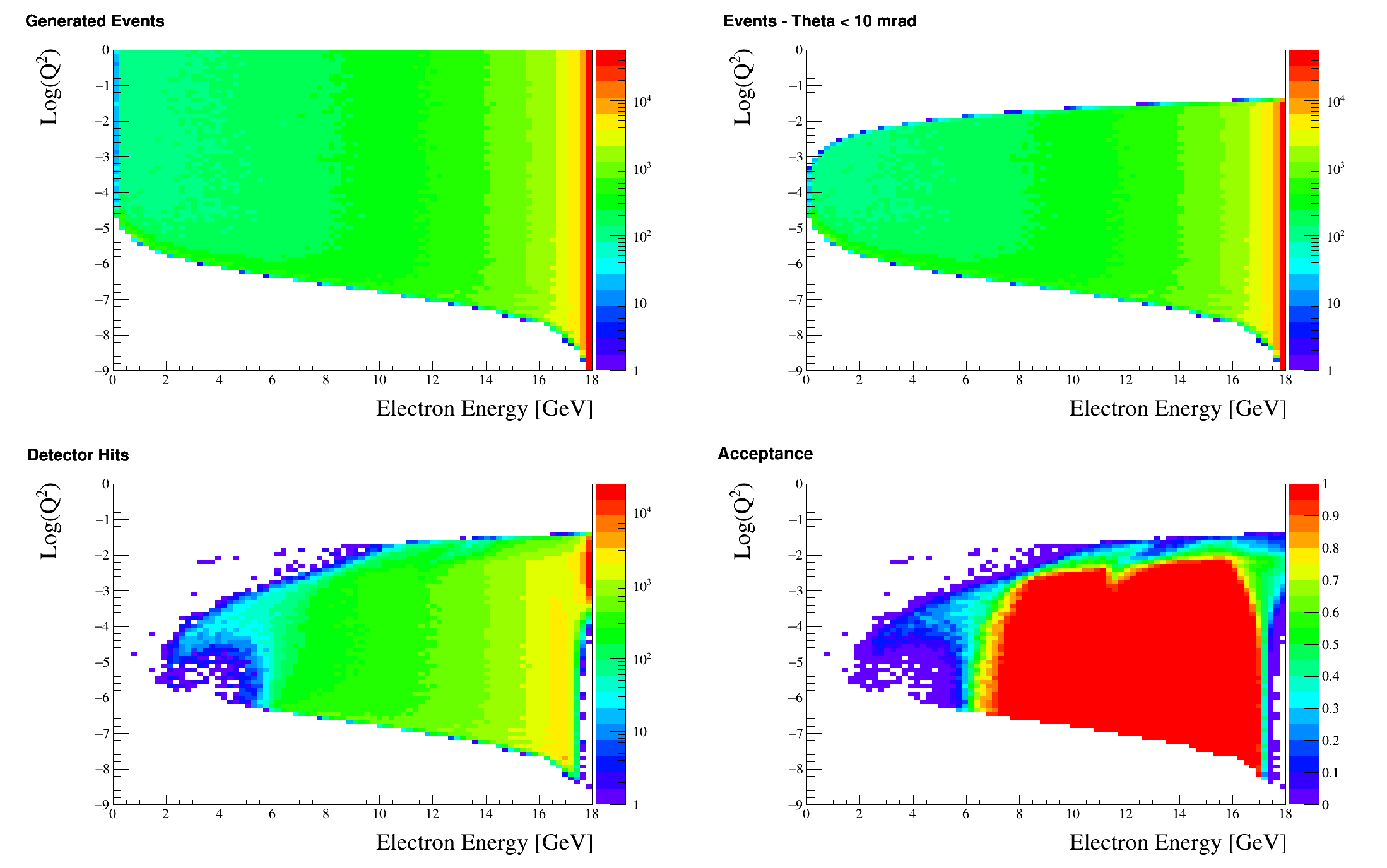}
}
\caption{\label{fig:acceptance} Generated and acceptance of events in terms of reaction Q$^2$ and scattered e$^-$ energy, from the quasi-real photoproduction event generator GETaLM\cite{ADAM2022108251}. Top Left: All generated events, Top Right: Events with electron polar angle <10 mrad, Bottom Left: Total hits in the taggers, Bottom Right: Acceptance of the high taggers given hits in 4 layers. }
\end{figure}

The acceptance of the detectors in this setup for a 18 GeV electron beam is shown in Figure \ref{fig:acceptance}.

\begin{figure}[H]
%\centering{\includegraphics[clip,width=\columnwidth]{EPICAcceptance_18GeV.png}
\centering{\includegraphics[clip,width=\columnwidth]{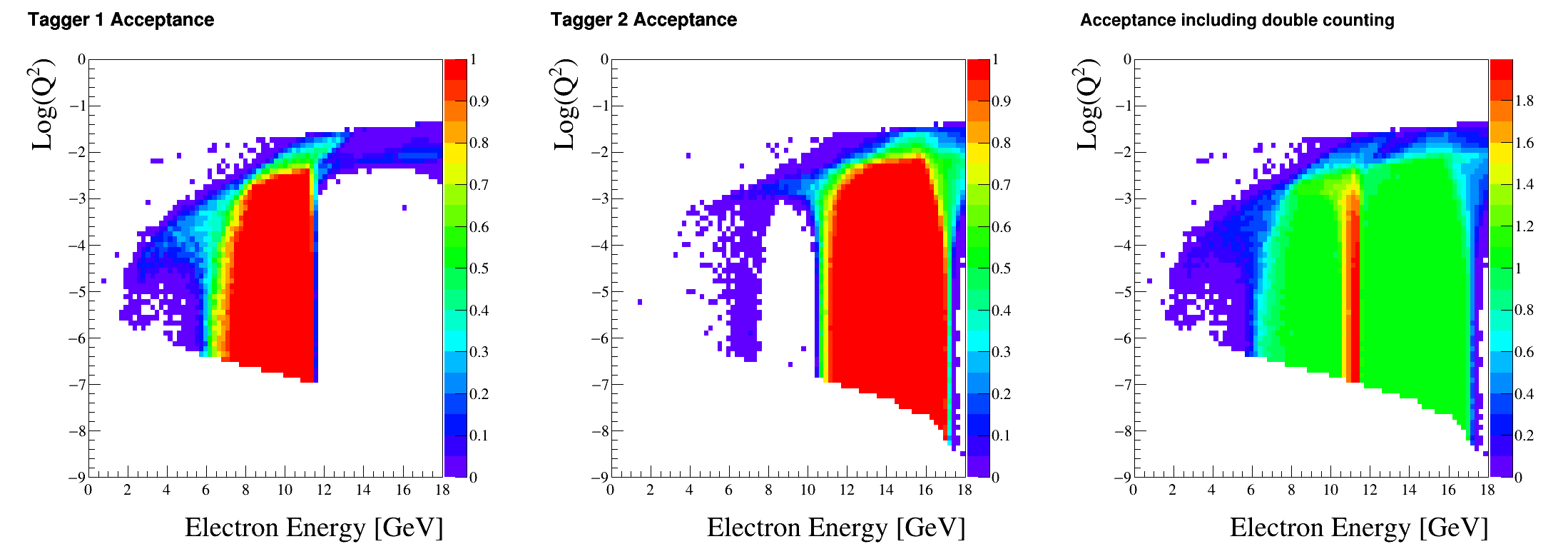}
}
\caption{\label{fig:acceptance2} Acceptance of the separate tagger modules in terms of reaction Q$^2$ and scattered e$^-$ energy. Left: Low energy tagger acceptance. Middle: High energy tagger acceptance. Right: Total acceptance including overlap.  }
\end{figure}

The acceptance and cross over of the two separate modules is shown in Figure \ref{fig:acceptance2}. An overlap between the taggers will be required to maximise acceptance and provide essential information for cross calibration. The current overlap in the simulation is 2 cm which is likely larger than will be required by the detector in the end. 

\subsubsection{\label{sec:reconstruction}Kinematic Reconstruction}
Due to the complexities created by the beam line magnetic optics, analytical approaches to reconstructing initial electron momenta are complicated and more computationally costly than necessary. Instead, here, machine learning techniques implemented with ROOT TMVA\cite{Hocker:2007ht} have been used to successfully reconstruct original electron kinematics from tracks in the detectors.

A dense deep neural network used reconstructed track parameters as inputs which pass through hidden layers before outputting the estimated generated electron kinematic variables. From the track position $\mathbf{P}$ and vector $\mathbf{V}$ four independent track parameters were selected for the input; x and y components of the unit vector and the y and z displacements where track intercepts the y-z plane at x=0,

\begin{equation}
    \left\{\mathbf{P^{x=0}_y}, \mathbf{P^{x=0}_z}, \mathbf{\hat{V}_x}, \mathbf{\hat{V}_y}\right\}.
\end{equation}

The selected targets for the output were electron energy E$_e$, $\theta_e$ and $\phi_e$, the track was always assumed to result from an electron, so the mass is fixed. The network training was guided by RMS errors between expected and predicted values. 
%In order to work with the azimuthal angle $\phi$, it was divided into components $\cos(\phi)$ and $\sin(\phi)$. 

\begin{equation}
    \left\{ E_e, \theta_e, \cos(\phi), \sin(\phi) \right\}
\end{equation}

\begin{figure}[H]
\centering{\includegraphics[clip,width=\columnwidth]{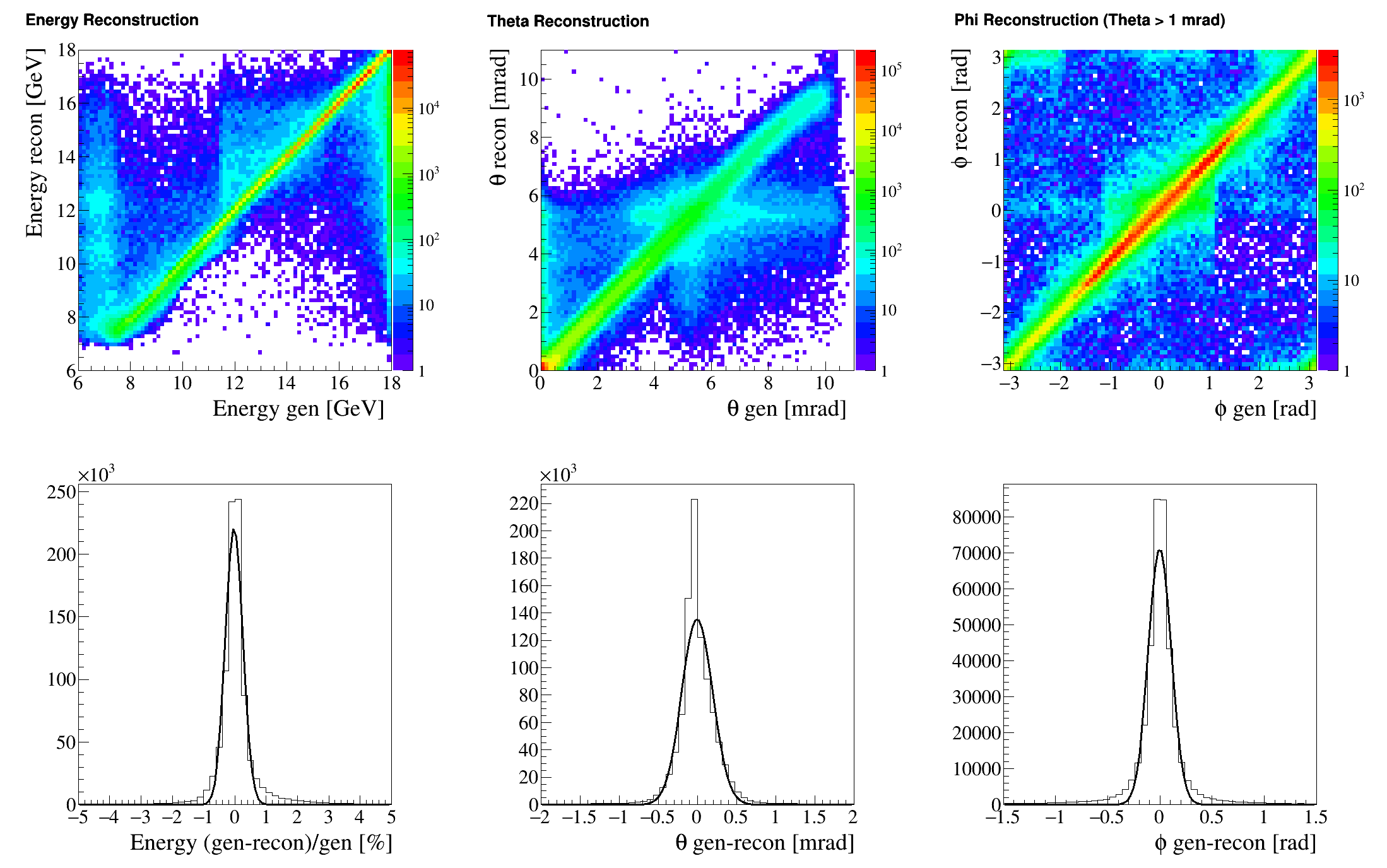}}
\caption{\label{fig:recon} Top - Reconstruction of the initial electron energy, $\theta$ and $\phi$ angles from fitted tracks. Bottom - Integrated reconstruction difference, the Gaussian fit $\sigma$ is presented in Table \ref{tab:resolutions}. }
\end{figure}

Further work on optimisation of network hyperparameters including the size and depth of the hidden layers still needs to be carried out with a view to reducing the computational demand without compromising the reconstruction resolution. Figure \ref{fig:recon} shows a representative reconstruction of the particle parameters from an in-vacuum tracker, the resulting resolutions are given in Table \ref{tab:resolutions}.

\begin{table}[H]
        \centering
    \begin{tabular}{c c c}
         Energy [\%] & $\theta$ [mrad] & $\phi$ [mrad]\\
         \hline 
         0.25 & 0.2 & 100\\
    \end{tabular}
    \caption{Integrated resolution of reconstructed particles.}
    \label{tab:resolutions}
\end{table}

\subsubsection{\label{sec:rates}Rates}

\begin{figure}[H]
\centering{\includegraphics[clip,width=\columnwidth]{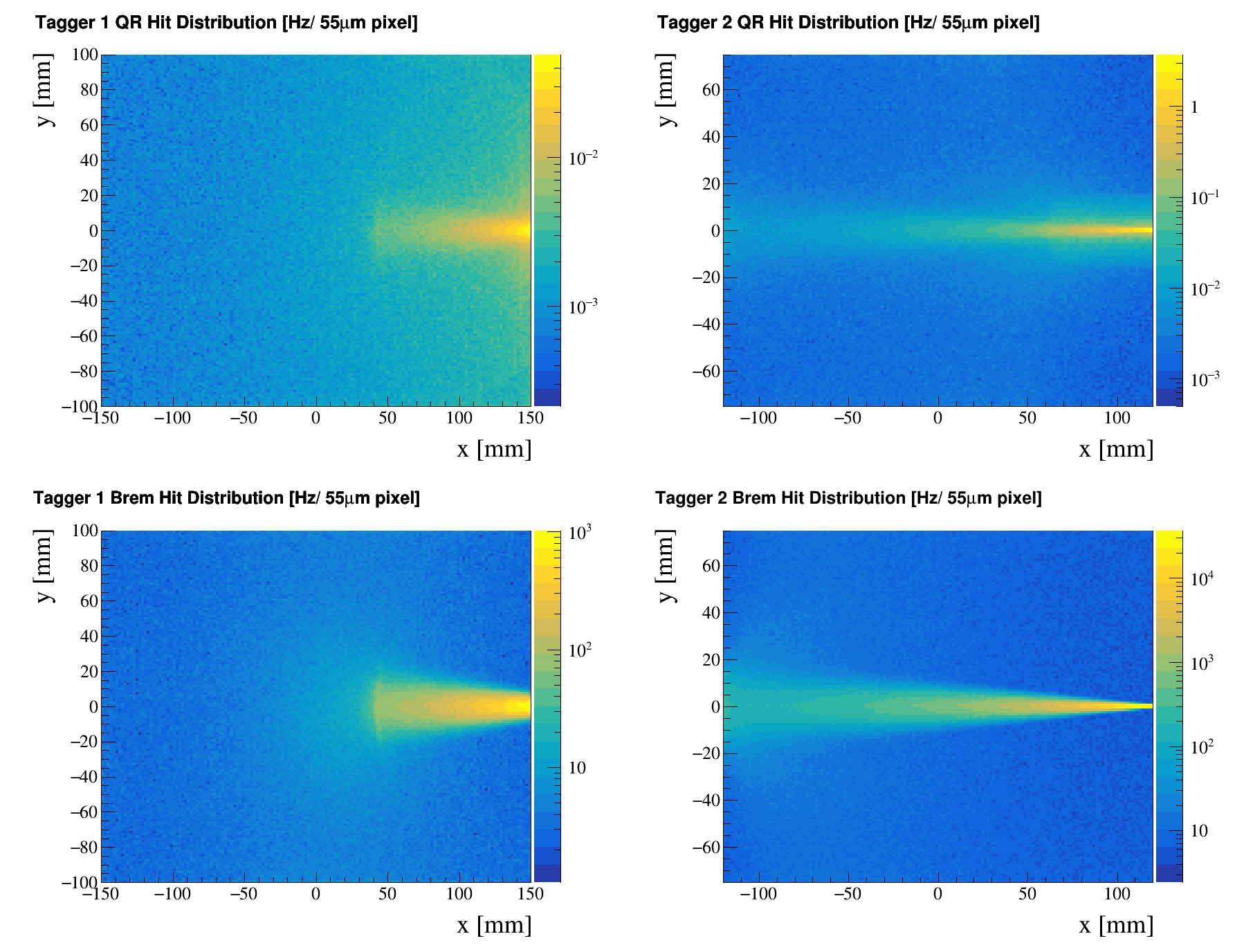}
\caption{\label{fig:rates} Rates per 55$\mu$m pixel on Taggers from Bremsstrahlung and Quasi-real events }
}
\end{figure}

The cross sections, integrated over the acceptance of the tracking detectors are 171 mb and 0.055 mb for Bremsstrahlung and Quasi-real (QR) events respectively, and the corresponding rates in the taggers are shown in Figure \ref{fig:rates}. It is clear that the majority of the hits on the pixel layers will be from bremsstrahlung electrons, and the challenge will be to select the electron corresponding to QR events associated with a hadron trigger in the main detector. It should also be noted that in the GEANT4 simulation, no consideration is made for charge spread, clustering, or multiple hits - ie an electron passing through a layer causes only a single pixel to fire, and estimates of cluster size are applied later when looking at specific pixel detectors. 

From the perspective of experimental design and data analysis, there are two separate quantities to consider: 
\begin{enumerate}
    \item The full rate of electrons passing through the trackers, which is about ten per bunch crossing, at a rate of 100MHz. \\This is the very high event rate that the combination of pixel detectors and readout will need to handle and keep in a buffer in wait for a corresponding trigger from the main hadron detector (500kHz). Only the 0.6\% of the buffered data which is in coincidence with a hadron trigger needs to be passed to to the main DAQ.
    \item The amount of data sent to the DAQ for each triggered event (at a rate of 500 kHz). \\In addition to the number of electrons per bunch (ten), this will depend on the details of the detector design (eg pixel size, number of layers, clustering capability).  
\end{enumerate} 

An estimate of these rates for a Timepix4 based tracker design is given in Section \ref{pixelcomp}.

%\begin{table}[H]
%    \centering
%    \begin{tabular}{l l l l}
%                          & \multicolumn{2}{ c }{Process} &           \\
%                          & Bremsstrahlung & Quasi-real & Ratio       \\
%        Cross section/event (mb)              & 171.3 & 0.055 & 3114  \\
%                                              &       &       &       \\ 
%        N(e$^-$)/bunch crossing               & 11.7   & 0.0037& 3162 \\
%        Tagger 1 hits/bunch crossing          & 1.5   & 0.00014& 10714\\
%        Tagger 2 hits/bunch crossing          & 7.6   & 0.00064& 11875\\
%                                              &       &       &       \\ 
%        N(e$^-$)/bunch crossing + Trig        & 11.7   & 0.74& 32     \\
%        Tagger 1 hits/bunch crossing + Trig   & 1.5   & 0.028& 107    \\
%        Tagger 2 hits/bunch crossing +Trig    & 7.6   & 0.128& 117    \\
%    \end{tabular}
%    \caption{Cross section and predicted event rate for generated sample 10 million electrons from the 18x275 beam %setting with a bunch luminosity of 0.068 mb$^{-1}$ and bunch crossing rate of 100 MHz. The final three rows (+ Trig %rates) consider only rates for events with a trigger generated by from a hadronic interaction (500 kHz)}
%    \label{tab:rates}
%\end{table}

%The rates observed in each tagger per 55$\mu$m pixel are shown in Figure\ref{fig:rates}.  

\section{\label{pixelcomp}Requirements for pixel detectors and readout}
The work presented here relates to Timepix Work Package within the UK's EIC Infrastructure Consortium \cite{stfcEIC}, where our remit is to investigate potential roles for Timepix hybrid pixel detectors at EIC. We quickly established that this technology could have most impact on the development of the Low Q$^{2}$ Tagger, and have focused on a Timepix4 design for that system. 

As described above, the Low Q$^{2}$ detector will consist of two separate trackers, based on pixel detectors. The number of layers, spacing and pixel size are still to be optimised on the basis of simulations, and will be determined in time for CD-2. The essential characteristics are angular resolution (since all other quantities are derived from polar and azimuthal angles), rate capability and background rejection. For pixel detectors, the angular resolutions relate to pixel size, or, more precisely, to the position resolution of the centroids of pixel clusters. From the simulations  it is clear that 55 $\mu$m pixels would provide very good resolution. Bigger pixels would still provide acceptable resolution, but high segmentation is even more important for rate capability, where the efficiency for separating multiple tracks within a single event needs to be as high as possible: For an electron-ion collision event  there are typically ten background bremsstrahlung electrons within the same beam bucket, each passing though all layers of a tagger and creating hits. Furthermore, in each layer there will be "singles" resulting from rescattering or synchrotron radiation together with hits from detector noise. However, we already have enough information to set some constraints on detector and readout technologies. We have used Timepix4 as the template for much of the development, and have had its dimensions, readout and rate capabilities as a strong influence in the development of the current design. However, we have tried, where possible, to used "generic" pixel detectors - particularly in the Geant4 simulations, with the aim of being able to evaluate other current, or emerging,  technologies, and we remain open to the possibility that a better, or more cost effective, solution might be the way forward. 

\begin{figure}[H]
\centering{
\includegraphics[clip,height=0.5\columnwidth]{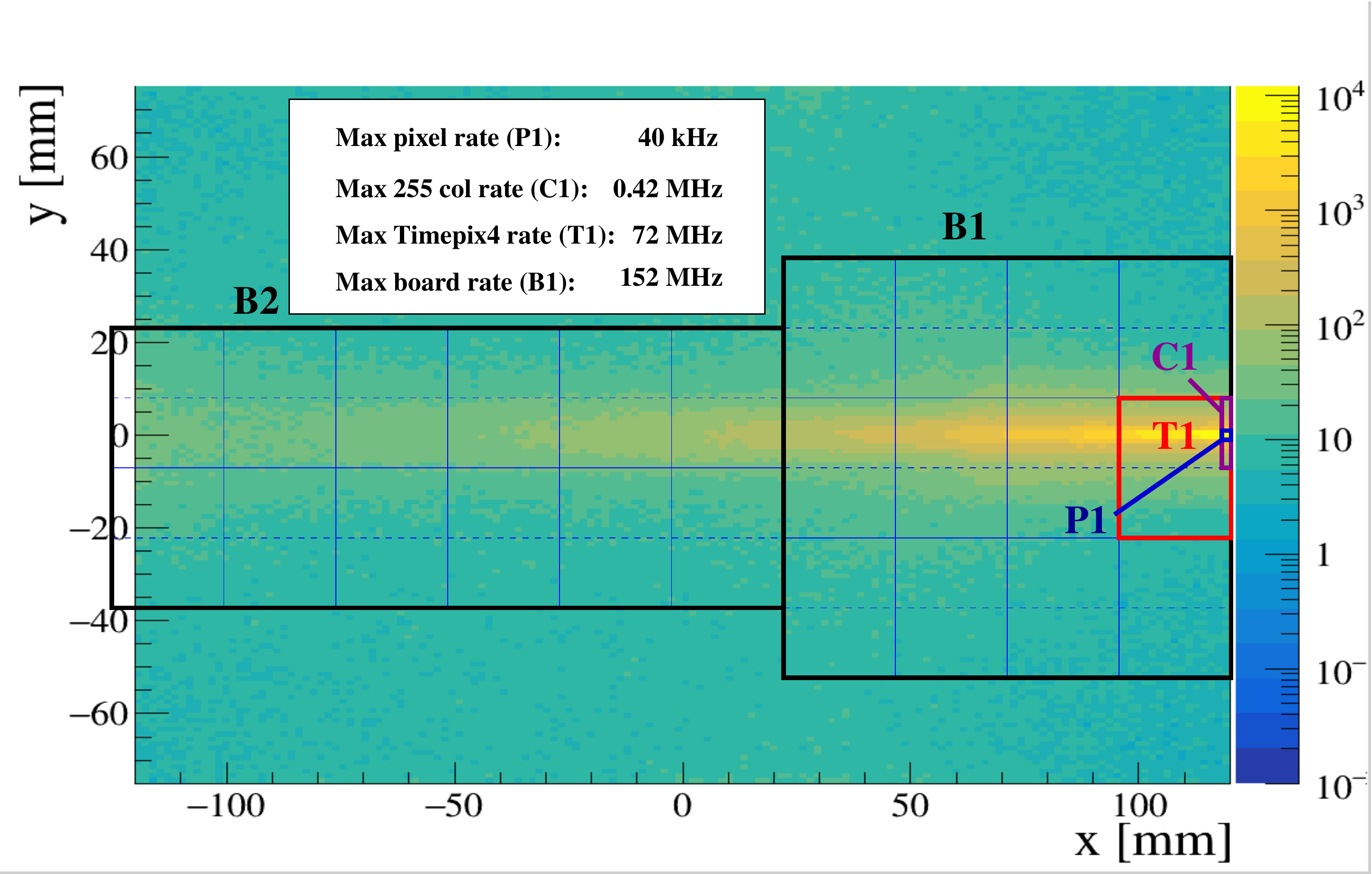}
}
\caption{\label{fig:tag2pixelrates} Maximum hit rates for bremsstrahlung electrons incident on Tagger 2, This example design is based on two board layers. The boards are outlined in black, and each contains twelve Timepix4 hybrids (outlined in blue)}
\end{figure}

It is already clear, both from a basic knowledge of the kinematics of bremsstrahlung and quasi-real events, and from preliminary simulations, that the intensity of electrons passing through the trackers will be distributed in a highly non-uniform way, with the bulk of the events close to the plane of the accelerator, and the flux increasing strongly towards the electron beamline. In particular, the rates on Tagger 2 are significantly higher than Tagger 1, with the hottest zone closest to the beamline (Figure \ref{fig:rates}). For an estimate of the relevant rates we focus on the bremsstrahlung distribution in Tagger 2 and superimpose a tracking layer geometry based on two boards, each consisting of 12 Timepix4 hybrids, as shown in Figure \ref{fig:tag2pixelrates}. The four Timepix4 detectors running across the centre of board 1 (B1) take the bulk of the events, with the very highest on the one closest to the electron beam (T1). The small vertical offset between the centre of the board and the accelerator plane is to ensure that the centre line (dashed), where the top and bottom vertical 255 pixel columns meet, does not coincide with the very high rate band. The maximum rate estimates which are significant for Timepix4, can be obtained by integrating over the relevant bins of the 2D histogram, and are shown in white inset on the figure. Although the dimensions of pixels and sensors are from Timepix4, these rates are \emph{detector agnostic}, in the sense that they merely quantify numbers of electrons passing though the 55$\mu$m pixels in the tracking plane. The average number of pixels registering a hit  above threshold (ie making a cluster) due to a passing electron will depend on detector specifics, and the number of "singles" due to synchrotron radiation is still unknown pending inclusion in the simulation.

We make a conservative estimate that each electron (or MIP) passing through a sensor produces a three pixel cluster, and assume the the large synchrotron related background contributes evenly across the layer at  a ratio of three times the layer maximum.  This results in the approximate rates shown in Table \ref{tab:maxrates}.

\begin{table}[H]
    \centering
    \begin{tabular}{lrl}
Maximum pixel rate                                                 & 120 & kHz \\
Maximum 255 column rate                                            & 121 & MHz\\
Maximum Timepix4 rate                                              & 214 & MHz\\
Maximum board (12 timepix4 sensors) rate                           & 460 & MHz\\ 
Maximum board (12 timepix4 sensors) rate, including synchrotron BG & 1.8 & GHz\\
Data readout per pixel                                             & 64 & bits\\
\textbf{Maximum board bit rate (64 x 1.8 GHz )}               & \textbf{115} & \textbf{Gb/s}
\end{tabular} 
\caption{\label{tab:maxrates} Maximum hit rates estimates based on bremsstrahlung electrons incident on Tagger 2, assuming 3 pixel cluster per MIP}
\end{table}

The rates in Table \ref{tab:maxrates} are based on number of pixels above threshold and define the desired capabilities of the pixel detector and associated readout. Each 12 sensor detector board would required a dedicated FPGA based server capable of handling the maximum board rate (115 Gb/s).

\begin{figure}[H]
\centering{
\includegraphics[clip,height=0.5\columnwidth]{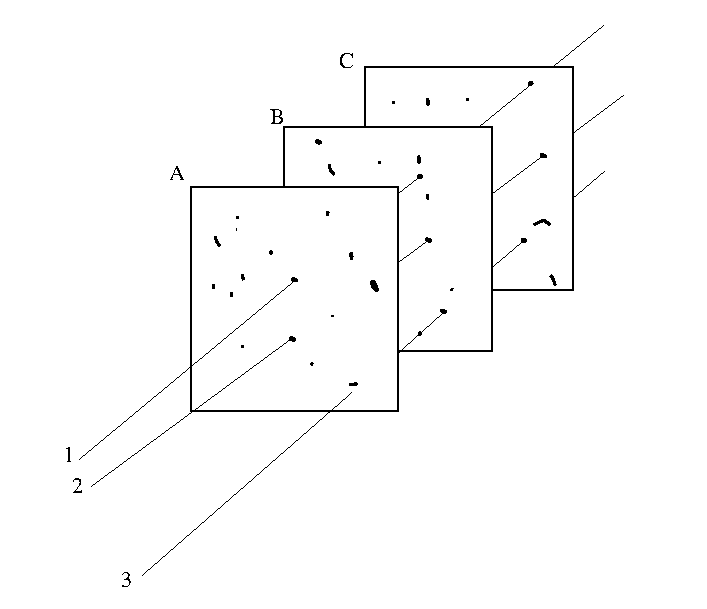}
}
\caption{\label{fig:recon3} Reconstructing events in a 3-layer tagger}
\end{figure}

The next step is to consider data reduction, buffering, how much information needs to be read out to identify tracks, and the rate at which data will be pushed to the main DAQ. Figure \ref{fig:recon3} illustrates the challenge. It represents a simplified version of the detector hits resulting from a single collision, where tracks 1 and 2 originate from the interaction region, and track 3 from somewhere else. Each of these passes though each of the layers, making "hits", which are characteristic of minimum ionising particles (MIPs). Then there are also other hits, as outlined above. In the ideal situation non-MIPs hits (shown as straggles, streaks, blotches and dots), would be eliminated at the layer level by a clustering algorithm, and this already makes some demands on the pixel detector/daq combination: The pixels must be small enough to make a cluster (ideally, using energy deposited to do weighted clustering), and the clustering analysis must be fast enough to deal with the event rate. We foresee the clustering being implemented in FPGA, and there are several published examples of this technique 
\cite{fpgaclustering1,fpgaclustering2}. 

\begin{figure}[H]
\centering{
\includegraphics[clip,height=0.3\columnwidth]{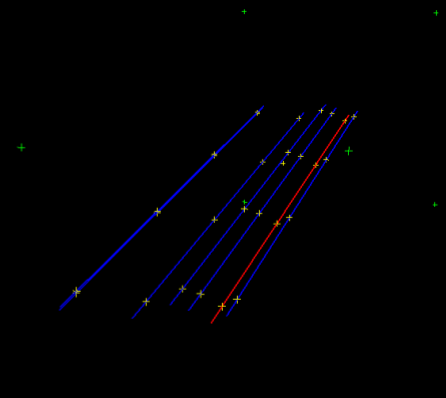}
}
\caption{\label{fig:jardatracks} Example of tracks reconstructed from simulated data with a 4 layer tracker. The bremsstahlung events are in blue, and the quasi-real event is in red. }
\end{figure}

Detailed simulation and analysis of the clustering and tracking have been carried out to establish the efficiency for identifying the quasi-real event within the bremsstrahlung background (Sec.\ref{sec:sim}). Events were generated using the ETaLM generator, and there are, on average, 19 bremsstrahlung electrons generated in every beam crossing, mostly striking one of the two trackers - with the bulk in Tagger 2, since it is closer to the electron beamline. Occasionally, there is also a quasi-real track, as illustrated in Figure \ref{fig:jardatracks}, where cluster centroids for four tracking layers are shown. A well established algorithm \cite{Duerdoth:1982hi} was used to  allocate a $\chi^{2}$ value to candidate tracks for each possible cluster combination. A tight cut on $\chi^{2}$ together with a cut to reject clusters closer the 0.5mm results in an efficiency of 95\% for identifying the quasi-real track. 

The following tables gives some estimates of rates based on current geant4 simulation with signal and background event generators, and detector segmentation that results in an average of 3 hits per MIP (ie pixel size of $\sim$55$\times$55 $\mu$m). 
\begin{table}[H]
    \centering
    \begin{tabular}{lrl}
Average number of electrons through tracker per bunch crossing & 10  & electrons\\
Total number of tracker layers (2 x 4)                         & 8   & layers\\
Total number of hits per bunch crossing (10 x 8)               & 80 & hits\\
Bits per cluster (x, y, time, energy, width: 5x2 bytes)        & 80  &  bits \\
Total bits per bunch crossing (80 x 80)                       & 6400 & bits\\
Hardon trigger rate                                            & 500   & kHz\\
\textbf{Total bit rate for hadron triggers} (6400 x 500 kHz) & \textbf{3.2} &  \textbf{Gb/s}\\
\textbf{Total bit rate, including random sample for BG} & \textbf{20} &  \textbf{Gb/s}\\
\end{tabular} 
\caption{\label{tab:daqrates} Estimated DAQ rate based on saving tracker cluster data for triggered events (3.2 Gb/s).  A random sample rate of about five times the trigger rate then results in an estimated data rate of 20Gb/s }\end{table}

As Tables \ref{tab:maxrates} and \ref{tab:daqrates} show, the solution proposed here results in a light load on the main experiment DAQ (3.2 Gb/s  is small compared to the rates from the central and forward detector systems), but makes high demands on the FPGA based readout boards. The design was driven by the decision to focus on the Timepix4 hybrid pixel detector, and take advantage of the recent developments in the corresponding readout systems and software development tools (Section \ref{timepix}). Any other candidate technologies would still need to provide the required resolutions and handle the rates outlined above, but might, for example, do less local processing and pass more data on to the main DAQ. If more than one technology is able to meet those fairly stringent requirements the ultimate choice will be based on other factors, such as cost, technology readiness, availability. Here we compare the requirements (estimated above) with the capabilities of Timepix4 and the two other technologies which have been proposed for other parts of the ePIC detectors: AC-LGAD \cite{lgad} and MAPS \cite{https://doi.org/10.48550/arxiv.2203.07626}. This information is still incomplete, and several parameters are not comparable. However, it has enabled us to conclude that timepix4 is the only feasible solution amongst the currently proposed technologies.

\begin{table}[H]
    \centering
    \begin{tabular}{l|r|r|r|r}
                              & Requirement             & Timepix4                 & AC-LGAD            &                  MAPS     \\
 \hline
Readout                       & -----                   & SPIDR4                   &  EICROC           &        Direct ?            \\
Pixel Size ($\mu$m)           &  50 $\times$ 50   & 55 $\times$ 55           &  500 $\times$ 500  &   20 $\times$ 20          \\
Sensor thickness ($\mu$m)    & -----                   &  100                      &  50                &   20                      \\
Detector size (pixels)        & -----                   & 512 $\times$ 448         &  64 $\times$ 64    &   Various                       \\
Detector area (cm$^{2}$)      & -----                   & 6.94                     &  10.24             &    Various                      \\   
Layer Area (cm$^{2}$)         & 100               & 83 (3x4 Timepix4)        &  92 (3x3)          &     Various                     \\
Power consumption (W/cm$^{2}$)& As low as possible      & 1.0                      &   0.4              &    0.15                   \\
                              &                         &                          &                    &                           \\        
Timing resolution (ns)        & < 12                    &  0.2                     &     0.03           &    9                       \\
Minimum threshold (fC)& -----                           & 1.2                      & 2.0                &    0.48                   \\
Individual pixel thresholds   & -----                   &  Yes                     &   Yes              &   No                       \\
Pixel hits in MIPS cluster    & -----                   & 3                        & 30                 & 5 ?                        \\
                              &                         &                          &                    &                            \\        
%%Max pixel rate (hits/pixel/s) & $\sim$ 6 $\times$ 10$^{3}$ & 10.8 $\times$ 10$^{3}$ &                 &                            \\
\textbf{Rate (various units)}      &                         &                         &                &                   \\
Hits/pixel/s (max)          &  120 $\times$ 10$^{3}$   &   > 10.8 $\times$ 10$^{3}$ (\textbf{Note 1})&      N/A   &                            \\
Hits/detector/s (max)       &  0.24 $\times$ 10$^{9}$ &   2.5 $\times$ 10$^{9}$ &            &                            \\
Bits/detector/s (max)       &   30 $\times$ 10$^{9}$ &   160 $\times$ 10$^{9}$ &              &                            \\
Bits/layer/s (integrated)   &  115 $\times$ 10$^{9}$ &  > 240 $\times$ 10$^{9}$              &                    &                           \\
                            &                        &                          &                    &                           \\ 
\end{tabular} 
\end{table}

\textbf{Note 1}: The \emph{Max Pix Rate} value of 10.8 kHz from the table in \ref{fig:timepixtable} is an average, based on the maximum readout rate of $\sim$ 2.8 MHz for a 256 pixel column. If the rate is distributed non-uniformly, as in the Low Q$^{2}$ Tagger, then the rate in high intensity pixels can go well beyond this - potentially as high as 1 MHz per pixel if the ASIC is configured a faster that standard charge collection, and some ToT precision is sacrificed. 

The requirements outlined in the table are mainly dictated by the location of the Low Q$^{2}$ Tagger: It is very close to the beam line and relies on measuring electrons with good angular resolution (Section \ref{sim}). It has to cope with the very high rates in that zone, particularly from the large bremsstrahlung background, and it must have the ability to separate out tracks which are close together within the same event (ie beam pulse). All the options listed can provide acceptable position resolution, and, although there is not data available to make a direct comparison in every category, it is clear that only the Timepix4 is able to handle the rates and sort out the tracks with good efficiency. 
MAPS has excellent position resolution and a very low material budget, optimised for multi-layer trackers, but would be unable to cope with the very high readout demands in the Far Backward region. AC-LGAD has a high readout bandwith and exceptional timing resolution, but the design based on large pixels and use of clusters for position resolution rules it out for this application, since the large number of overlapping clusters would make it almost impossible to separate the multiple tracks in an event, and, additionally, the number of pixels firing in a layer for an event would be much higher due to the intrinsic design of the AC-LGAD, where many pixels are needed to provide the position resolution. 

In summary, we believe Timepix4 is close to being the optimum solution available from the current detector technologies. Furthermore, Timepix4, together with SPIDR4 are the most mature of the technologies discussed, and already close to being \emph{off-the-shelf} components. It is also worth noting that Timepix4 is a hybrid sensor, opening up the option of future upgrades to the design. For example, replacement of the current, standard, pixellated silicon sensor with a Inverse LGAD \cite{Curr_s_2020} could potentially improve the timing resolution to 10s of picoseconds without and need to redesign any other aspects of the readout boards or infrastructure.

\section{\label{design}Preliminary design}
\begin{figure}[H]
\centering{
\includegraphics[clip,height=0.5\columnwidth]{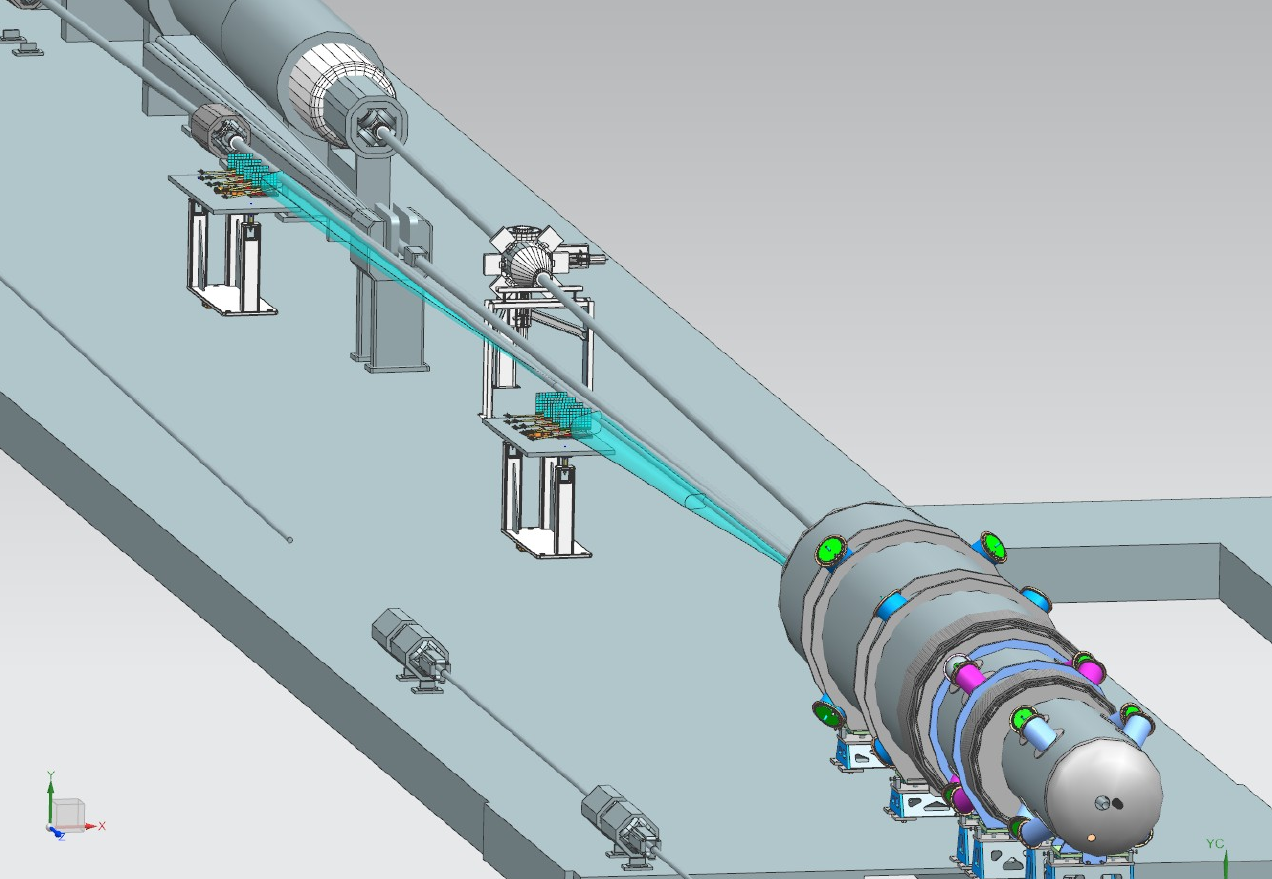}
}
\caption{\label{fig:beamlinecad} Locations of the Low Q$^2$ Tagger trackers in the EIC CAD}
\end{figure}

\begin{figure}[H]
\centering{
\includegraphics[clip,height=0.5\columnwidth]{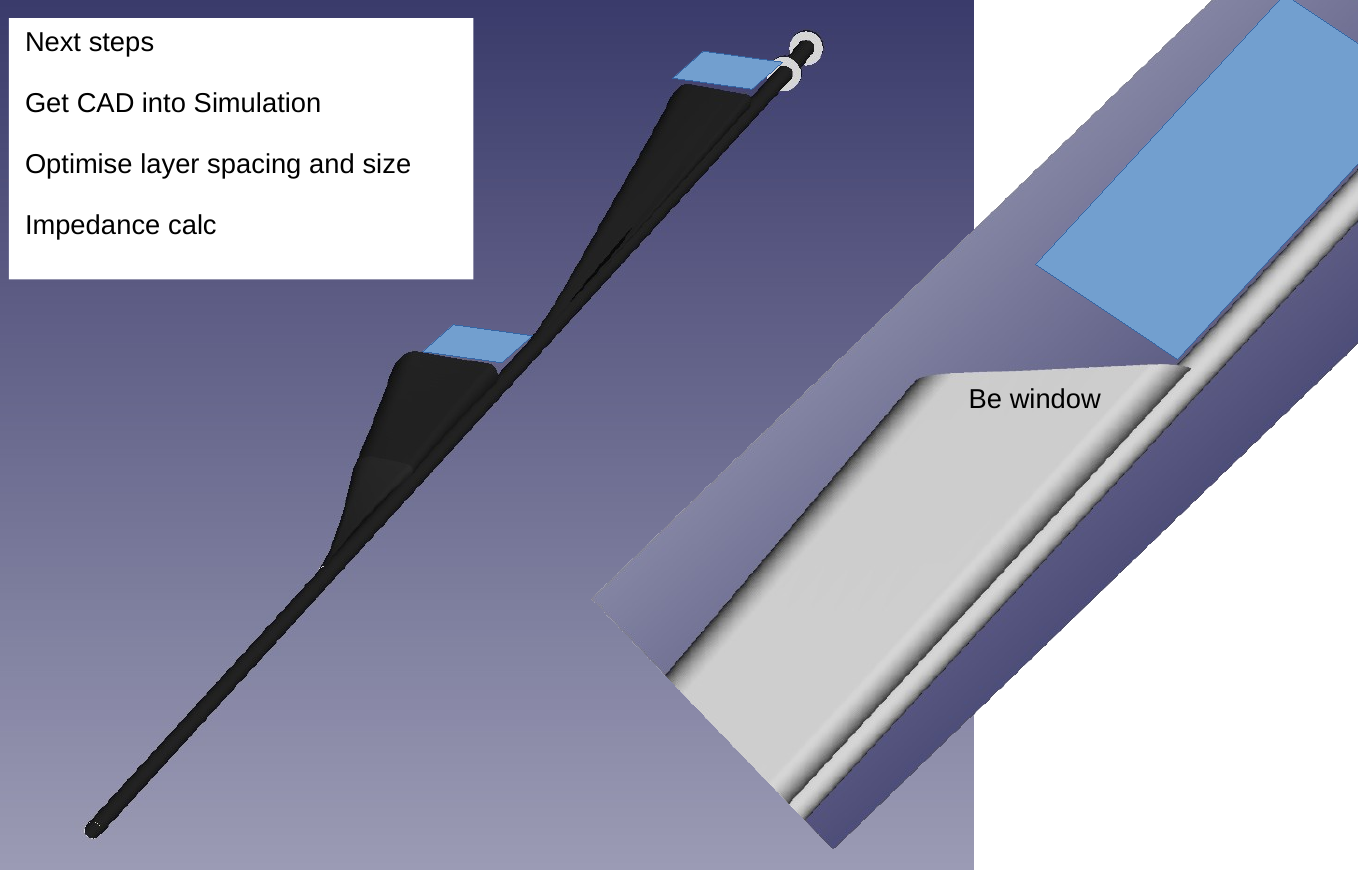}
}
\caption{\label{fig:vacboxcad} Low Q$^2$ Vacuum box and window structure in the EIC CAD}
\end{figure}

\begin{figure}[H]
\centering{
\includegraphics[clip,height=0.5\columnwidth]{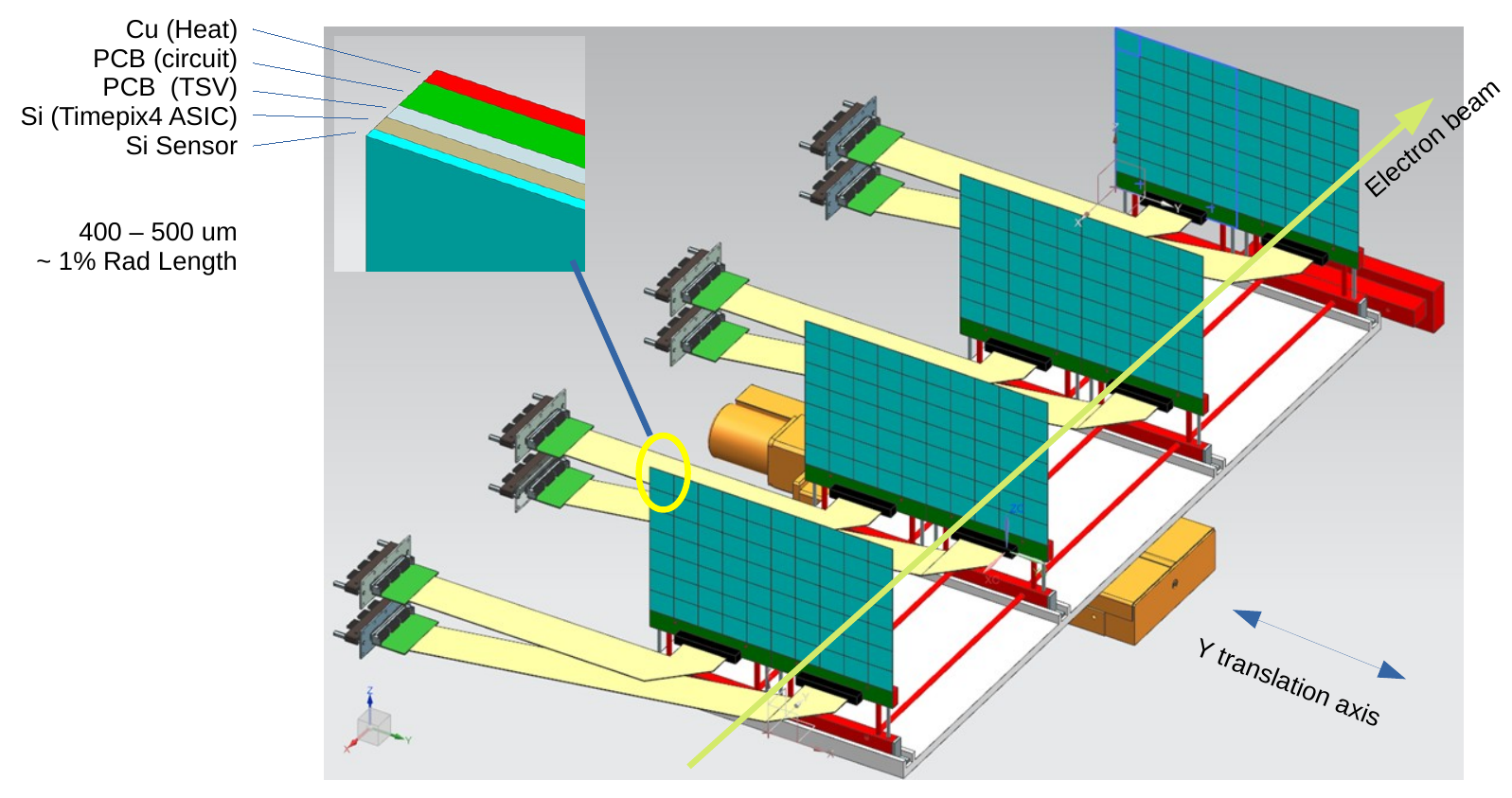}
}
\caption{\label{fig:trackerscad} Low Q$^2$ Tagger trackers in the EIC/ePIC CAD model}
\end{figure}

A preliminary low Q$^{2}$ detector based on two Timepix4 trackers, each with 4 layers is now part of the baseline design for ePIC. The current CAD model has an in-beam solution, where the trackers are housed in the main electron beamline vacuum box. This is shown in Figures \ref{fig:beamlinecad} and \ref{fig:vacboxcad}. A schematic arrangement of the frame houling the four layers of Timepix4 trackers is also shown in Figure \ref{fig:trackerscad}. This section will be updated as the design progresses.

\section{\label{conclusion}Conclusion and outlook}

A strong physics case has been made for the inclusion of a Low Q$^2$ tagger in the Far Backward region of the ePIC detector at EIC. Simulation and rate calculations have demonstrated that a Tagger based on Timepix4 tracking detectors is the best solution based on current technologies, and this is now included in the baseline design. 

A funding application for the construction of Low Q$^2$ tagger has been submitted to UKRI as part of a bigger EIC infrastructure bid. Further development of the design will continue as this progresses, and this document will be updated in the future to reflect that.
\renewcommand*{\bibfont}{\raggedright}
\bibliographystyle{bry2}
\bibliography{timepixRefs}

\end{document}